\documentclass{article}
\usepackage[utf8]{inputenc}
\usepackage{amsmath}
\usepackage{graphicx}
\usepackage{adjustbox}
\usepackage{url}
\usepackage{hhline}
\usepackage{float,subfigure}
\usepackage{authblk}
\restylefloat{table}
\graphicspath{ {./images/} }
\usepackage{lineno}
\usepackage{placeins}

\title{Optimal renormalization of multi-scale systems}
\author[b,1]{Jacob Price}
\author[c]{Brek Meuris}
\author[b,a]{Madelyn Shapiro}
\author[a,d]{Panos Stinis}

\affil[a]{Pacific Northwest National Laboratory, Richland, WA 99354, USA}
\affil[b]{Department of Mathematics and Computer Science, University of Puget Sound, Seattle, WA 98416, USA}
\affil[c]{Department of Mechanical Engineering, University of Washington, Seattle, WA 98195, USA}
\affil[d]{Department of Applied Mathematics, University of Washington, Seattle, WA 98195, USA}

\begin{document}
\maketitle
\begin{abstract}
While model order reduction is a promising approach in dealing with multi-scale time-dependent systems that are too large or too expensive to simulate for long times, the resulting reduced order models can suffer from instabilities. We have recently developed a time-dependent renormalization approach to stabilize such reduced models. In the current work, we extend this framework by introducing a parameter that controls the time-decay of the memory of such models and optimally selecting this parameter based on limited fully resolved simulations. First, we demonstrate our framework on the inviscid Burgers equation whose solution develops a finite-time singularity. Our renormalized reduced order models are stable and accurate for long times while using for their calibration only data from a full order simulation before the occurrence of the singularity. Furthermore, we apply this framework to the 3D Euler equations of incompressible fluid flow, where the problem of finite-time singularity formation is still open and where brute force simulation is only feasible for short times. Our approach allows us to obtain for the first time a perturbatively renormalizable model which is stable for long times and includes all the complex effects present in the 3D Euler dynamics. We find that, in each application, the renormalization coefficients display algebraic decay with increasing resolution, and that the parameter which controls the time-decay of the memory is problem-dependent.
\end{abstract}

\section{Introduction}
Real-world applications from molecular dynamics to fluid turbulence and general relativity can give rise to systems of differential equations with tremendous numbers of degrees of freedom. More often than not, these systems are multi-scale in nature, meaning that the evolution of the various degrees of freedom covers a large range of spatial and temporal scales. When the degrees of freedom can be simply sorted into a few discrete collections of scales, a variety of techniques allow for simulation and analysis ( see e.g. \cite{givon2004extracting}).

However, there are many cases that lack this clear scale separation. For example, when a generic  partial differential equation is converted to a system of differential equations through the Fourier transform, the resulting system is comprised of degrees of freedom whose scales vary continuously with no clear demarcation between ``fast'' and ``slow.'' Furthermore, many multi-scale systems include a prohibitively large number of degrees of freedom, such that the simulation of all of them is impossible. Through reduced order modeling, we seek to construct a related system of differential equations for a subset of the full degrees of freedom whose dynamics accurately approximate the dynamics of those degrees of freedom in the full system. One mathematical framework for constructing reduced order models (ROMs) is the Mori-Zwanzig formalism (MZ). Originally developed in the context of statistical mechanics \cite{zwanzig1961memory}, the formalism has been modernized as a mathematical tool \cite{chorin2002optimal,chorin2007problem}. This formalism allows one to decompose the dynamics of a subset of variables (the \emph{resolved} variables) in terms of a Markov term, a noise term, and a memory integral. This decomposition elucidates the interaction between the resolved variables and the rest of the variables, called \emph{unresolved}. Based on various approximations, this framework has led to successful ROMs for a host of systems (see e.g \cite{chorin2002optimal, bernstein2007optimal,stinis2012numerical,li2015memory,lei2016,parish2017}). Except for special cases, it is difficult to guarantee that the reduced models will remain stable. We have developed a time-dependent version of the \emph{renormalization} concept from physics \cite{goldenfeld1992,georgi1993}, in which we attach time-dependent coefficients to the memory terms in the ROM. This has led to success in stabilizing reduced models of this type \cite{stinis2013renormalized,stinis2015renormalized,price2019renormalized}.

The MZ formalism has been previously used to develop ROMs for Burgers and 3D Euler \cite{hald2007optimal,stinis2007higher,stinis2013renormalized,stinis2015renormalized}. Those models were based on approximations of the memory term which assume various degrees of ``long memory'' i.e., the assumption that the unresolved variables evolve on timescales that are comparable to the resolved variables. 
Such an assumption is appropriate for inviscid Burgers and 3D Euler equations (and high Reynolds number fluid flows in general), given the vast range of active scales present in the solution. Recently \cite{price2019renormalized}, we have developed a novel expansion of the memory term (dubbed the ``complete memory approximation'') which also assumes long memory but avoids other simplifying approximations. In the current work, we introduce a parameter that allows to control the time-decay of the memory and can be selected based on limited fully resolved simulations (Section 1). 
We apply this to the inviscid Burgers equation to demonstrate the stability and accuracy of the optimized renormalized ROMs (Section 2). We then present results for renormalized ROMs of the 3D Euler equations (Section 3). To the best of our knowledge this is the first time-dependent perturbative renormalization approach for 3D Euler which includes all the complexity of the Euler dynamics.

\section{The complete memory approximation of MZ}

Previous work \cite{price2019renormalized} includes a comprehensive overview of the MZ formalism and the construction of ROMs from it by way of the complete memory approximation (CMA). Here we present an abridged version. Consider a system of autonomous ordinary differential equations (ODEs)
$\frac{d \mathbf{u}(t)}{dt} = \mathbf{R}(\mathbf{u})$ 
augmented with an initial condition $\mathbf{u}(0)=\mathbf{u}^0$. Let $\mathbf{u}(t) = \{u_k(t)\}$, $k\in F\cup G$. We separate $\mathbf{u}(t)$ into resolved variables $\hat{\mathbf{u}}=\{u_i(t)\}$, $i\in F$ and unresolved variables $\tilde{\mathbf{u}}=\{u_j(t)\}$, $j\in G$ where $F$ and $G$ are disjoint. Let $R_k(\mathbf{u})$ be the $k$th entry in the vector-valued function $\mathbf{R}(\mathbf{u})$. We can transform this nonlinear system of ODEs into a linear system of PDEs by way of the Liouvillian operator \cite{chorin2000optimal,chorin2002optimal}:
$
\mathcal{L}=\sum_{k\in F\cup G} R_k(\mathbf{u}^0)\frac{\partial}{\partial u_k^0}.\label{L}
$

It can be shown that
\begin{equation}
\frac{du_k(t)}{dt} = \frac{\partial}{\partial t}e^{t\mathcal{L}}u_k^0 = e^{t\mathcal{L}}\mathcal{L}u_k^0.\label{semigroup}
\end{equation} 
Consider the space of functions that depend upon $\mathbf{u}^0$. Let $P$ be an orthogonal projection onto the subspace of functions depending only on the resolved variables $\hat{\mathbf{u}}^0$. For example, $Pf$ might be the conditional expectation of $f$ given the resolved variables and an assumed joint density. Let $Q=I-P$. Then, we can decompose the evolution operator $e^{t\mathcal{L}}$ using Dyson's formula into:
\begin{equation}
\frac{du_k}{dt} = e^{t\mathcal{L}}P\mathcal{L}u_k^0 +e^{tQ\mathcal{L}}Q\mathcal{L}u_k^0 + \int_0^t e^{(t-s)\mathcal{L}}P\mathcal{L}e^{sQ\mathcal{L}}Q\mathcal{L}u_k^0\,\mathrm{d}s.\label{MZ}
\end{equation}This is the Mori-Zwanzig identity. It is simply a rewritten version of the original dynamics. The first term on the right hand side in (\ref{MZ}) is called the Markov term, because it depends only on the instantaneous values of the resolved variables. The second term is called `noise' and the third is called `memory'. We again project the dynamics (the noise term vanishes):
\begin{equation}
\frac{dPu_k}{dt} = Pe^{t\mathcal{L}}P\mathcal{L}u_k^0+P\int_0^te^{(t-s)\mathcal{L}}P\mathcal{L}e^{sQ\mathcal{L}}Q\mathcal{L}u_k^0\,\mathrm{d}s.\label{reduced_MZ}
\end{equation} For $k\in F$, (\ref{reduced_MZ}) describes the projected dynamics of the resolved variables. It gives the average behavior of $u_k$. The system is not closed, however, due to the presence of the orthogonal dynamics operator $e^{sQ\mathcal{L}}$ in the memory term. In order to simulate the dynamics of (\ref{reduced_MZ}) exactly, one needs to evaluate the second term which requires the dynamics of the unresolved variables. Dropping the memory term and simulating only the Markov term may not accurately reflect the dynamics of the resolved variables in the full simulation. Any multi-scale dynamical model must approximate or compute the memory term, or argue convincingly why the memory term is negligible. In a previous work, it was shown that even when the memory term is small in magnitude, neglecting it leads to inaccurate simulations \cite{price2019renormalized}.

Define the Markov term as
$
R_k^0(\hat{\mathbf{u}}) = Pe^{t\mathcal{L}}P\mathcal{L}u_k^0,
$
and the memory term as
$
\mathcal{M}_k = P\int_0^te^{(t-s)\mathcal{L}}P\mathcal{L}e^{sQ\mathcal{L}}Q\mathcal{L}u_k^0\,\mathrm{d}s.
$
The simplest possible approximation of the memory integral is to assume the integrand is constant. In this case, the memory integral becomes:
$
\mathcal{M}_k \approx tPe^{t\mathcal{L}}P\mathcal{L}Q\mathcal{L}u_k^0.
$
This model is called the $t$-model and it has been used to successfully construct reduced order models for a variety of problems \cite{chorin2002optimal,stinis2012numerical,chorin2007problem,hald2007optimal,bernstein2007optimal,chandy2009t}. 

The CMA improves upon the accuracy of the $t$-model by constructing a series representation of $\mathcal{M}_k$ in powers of $t$. We begin by rewriting the memory term using the Taylor expansions of $e^{-s\mathcal{L}}$ and $e^{sQ\mathcal{L}}$ and then computing the integral termwise (assuming the integrand is sufficiently smooth that this interchange of integral and sum is valid):
\begin{equation}
\mathcal{M}_k
=Pe^{t\mathcal{L}}\left(\sum_{i=0}^\infty \sum_{j=0}^\infty \frac{(-1)^it^{i+j+1}}{i!j!(i+j+1)}\mathcal{L}^iP\mathcal{L}(Q\mathcal{L})^{j}Q\mathcal{L}u_k^0\right).\label{eq:novel}
\end{equation}We arrange the terms by powers of $t.$ Note that this arrangement implies long memory since it assumes absence of timescale separation between $e^{t\mathcal{L}}$ and $e^{tQ\mathcal{L}}.$ We find\small
\begin{align}
\mathcal{M}_k =&tPe^{t\mathcal{L}}P\mathcal{L}Q\mathcal{L}u_k^0 - \frac{t^2}{2}Pe^{t\mathcal{L}}\left[\mathcal{L}P\mathcal{L}Q\mathcal{L}-P\mathcal{L}Q\mathcal{L}Q\mathcal{L}\right]u_k^0+O(t^3). \label{unclosed}
\end{align}\normalsize The $O(t)$ term is the $t$-model once again. The $O(t^2)$ term, presents a new problem. The expression $\mathcal{L}P\mathcal{L}Q\mathcal{L}u_k^0$ is not projected onto the resolved variables prior to its evolution. It is a function of all modes, not just the resolved ones. This makes it impossible to compute as part of a reduced order model except in very special cases. 

To close the model in the resolved variables we construct an additional reduced order model for the problem term. First, note that
$
Pe^{t\mathcal{L}}\mathcal{L}P\mathcal{L}Q\mathcal{L}u_k^0=\frac{\partial}{\partial t} Pe^{t\mathcal{L}}P\mathcal{L}Q\mathcal{L}u_k^0.
$
That is, it is itself a time derivative. Second, apply MZ with the CMA construction for the evolution of $Pe^{t\mathcal{L}}P\mathcal{L}Q\mathcal{L}u_k^0.$ We find that only the Markov term of the MZ-CMA model for $Pe^{t\mathcal{L}}P\mathcal{L}Q\mathcal{L}u_k^0$ contributes at the $O(t^2)$ level.  \eqref{unclosed} becomes 
\begin{align}
\mathcal{M}_k =&tPe^{t\mathcal{L}}P\mathcal{L}Q\mathcal{L}u_k^0 - \frac{t^2}{2}Pe^{t\mathcal{L}}P\mathcal{L}\left[P\mathcal{L}-Q\mathcal{L}\right]Q\mathcal{L}u_k^0+O(t^3).\label{closed}
\end{align}
where all the terms are now projected prior to evolution and so involve {\it only} resolved variables. 

Many of the higher order terms in \eqref{eq:novel} contain a leading $\mathcal{L}$ like the ``problem term'' in $O(t^2)$ discussed above. In each case, we can construct a ROM for the problem term and approximate it by expanding the memory term in a series. The terms in this series will also include leading $\mathcal{L}$ terms, but we can repeat our procedure indefinitely. In this manner, we can construct an approximation for \eqref{eq:novel} in which every term has a leading $P$ before the evolution operator is applied. The resulting series is written as:
$
\mathcal{M}_k = \sum_{i=1}^\infty \frac{(-1)^{i+1}t^i}{i!} R_k^i(\hat{\mathbf{u}}).
$
We can uniquely define $R_k^i(\hat{\mathbf{u}})$ for any positive integer $i$. We automated this process in a symbolic notebook, which is available in \cite{MZrepos}. Different approximation schemes can be constructed by truncating this series at different orders of $t.$

The resulting ROMs can be unstable. In earlier work,  for a simpler memory approximation than CMA, it was found that \emph{renormalization} rendered the reduced order models for Euler's equations stable \cite{stinis2013renormalized,stinis2015renormalized}. We attach additional coefficients to each term in the series, such that the terms represent an \emph{effective memory}, given knowledge only of the resolved modes \cite{georgi1993}. The evolution equation for a reduced variable becomes
\begin{equation}
\frac{dPu_k}{dt} = R_k^0(\hat{\mathbf{u}}) + \sum_{i=1}^n \alpha_i(t)t^i R_k^i(\hat{\mathbf{u}}).\label{renormalized}
\end{equation}Here,we allow the renormalization coefficients $\alpha_i(t)$ to be time dependent. This gives us the flexibility to allow the functional form of the effective memory to be dynamic if necessary. In effect, this dictates the length of the memory. These coefficients must be chosen in a way that captures information we know about the memory term.

For the examples we have examined, if we do not renormalize so that $\alpha_i(t)$ are given by $\frac{(-1)^{i+1}}{i!}$, we find that the simulations are unstable for all except the $t$-model alone. In \cite{stinis2013renormalized,stinis2015renormalized}, ROMs for inviscid Burgers and 3D Euler were stabilized with constant renormalization coefficients ($\alpha_i(t)=b_i\frac{(-1)^{i+1}}{i!}$, where $b_i \neq 1$),  and in \cite{price2019renormalized},  ROMs of the Korteweg-de Vries equation  were stabilized with algebraically decaying renormalization coefficients $\alpha_i(t) = a_it^{-i}$ (where the $\frac{(-1)^{i+1}}{i!}$ have been absorbed in the $a_i$). In the current work, we posit that the renormalization coefficients $\alpha_i(t)$ are given by:
 \begin{equation}
     \alpha_i(t) = a_it^{-i\tau}.
 \end{equation}The parameter $\tau$ allows more flexibility and characterizes the rapidity of the memory decay. 

\section{Renormalized ROMs for the inviscid Burgers equation}\label{burgers2Deuler}

Consider the inviscid 1D Burgers equation,
\begin{equation}
u_t + uu_x = 0,
\end{equation}on a periodic domain with $u(x,0) = \sin(x)$. This problem produces a shock at time $T=1.$ Once the shock forms, it dominates the dynamics of the system. In previous work, renormalized ROMs that approximate the memory term differently than the CMA were used to approximate this system \cite{stinis2013renormalized,stinis2015renormalized}. We revisit this problem now with the CMA with dynamic renormalization. Let $u(x,t) = \sum_{k\in F\cup G} u_k(t) e^{ikx}$where $F=\left[-N+1,\dots, N-1\right]$ and $G = \left[ -M+1,\dots,-N,N,\dots, M-1\right]$ for $N<M$. Let $\mathbf{u} =\{u_k(t)\}_{k\in F\cup G}$, and let $\hat{\mathbf{u}} = \{u_k\}_{k\in F}$ and $\tilde{\mathbf{u}}=\{u_k\}_{k\in G}$. The equation of motion for the Fourier mode $u_k$ is
\begin{equation}
\frac{d u_k}{dt} =R_k(\mathbf{u}) = -\frac{ik}{2}\sum_{\substack{p+q=k\\p,q\in F\cup G}}u_p u_q.\label{full_Burgers}
\end{equation}
 We define $\mathcal{L}$ as in \eqref{L} such that $\mathcal{L}u_k^0 = R_k(\mathbf{u}^0)$ and the projector $P$ as $Pf(\mathbf{u}^0)=Pf(\hat{\mathbf{u}}^0,\tilde{\mathbf{u}}^0)=f(\hat{\mathbf{u}}^0,0)$.  Note that the initial condition lies entirely in the projected domain. With these definitions, we can compute the memory terms $R_k^i(\hat{\mathbf{u}})$ from Eq. (\ref{renormalized}). For the case of Burgers but also 3D Euler in the next section, we have chosen $N=M/2$ due to the quadratic nonlinearity. Other choices will be explored elsewhere.

With the exception of the $t$-model, the resulting unrenormalized reduced order models are not stable. We will use the rates of change of the energy in each resolved mode $E_{k}(t) = |u_{k}|^2$ as the quantities we attempt to match in the renormalization process. This choice is reasonable because it is known that energy moves from low-frequency modes to high-frequency modes as the shock develops, but that the Markov term is incapable of capturing this since it conserves energy in the resolved modes. The rate of change of the energy in a particular mode in the full model is:
$
\Delta E_{k}(t) = R_{k}(u)\overline{u}_{k}+u_{k}\overline{R}_{k}(u).
$
In a reduced order model, each term in the series has its own contribution to the energy derivative:
$
\Delta E_{k}^i(t) = R^i_{k}(\hat{u})\overline{u}_{k}+u_{k}\overline{R}^i_{k}(\hat{u}).
$

In order to compute the renormalized coefficients for the ROM we need to collect data from the full order system while it is still well resolved. At each timestep, we calculate the rate of change of energy flowing out of $F = \{\mathbf{k}\;\;|\;\;\mathbf{k}\in[-M/2+1,M/2-1]\}$ by computing $\frac{1}{2}\sum_{\mathbf{k}\in F}t  \Delta E_k^1(t).$ We restrict ourselves to timesteps $t^*$ where this quantity is less than $10^{-10}$. In other words, we use the $t$-model memory term to monitor the transfer of energy out of the first $M/2$ wavenumbers of the solution of the full system. In previous work, it has been shown that this can provide a reliable, if conservative, estimate of the transfer of energy across wavenumbers \cite{stinis2009phase}. Alternatively, one can monitor the transfer of energy using the whole dynamics i.e.  $\frac{1}{2}\sum_{\mathbf{k}\in F}  \Delta E_k(t),$ although this does not change the estimated renormalized coefficients. 

Consider a reduced order model of resolution $N$ that includes CMA terms up through order $n$. Using these $t^*$ steps, we minimize Eq. (\ref{scaling_cost}) for the pre-factors $a_i.$

\small
\begin{equation}
C_{N,n}(\mathbf{a},\tau) = \sum_{k\in F}\sum_{t\in t^*} \left(\Delta E_{k} - \Delta E_{k}^0-\sum_{i=1}^{n}a_it^{-i\tau}t^i \Delta E_{k}^i\right)^2,\label{scaling_cost}
\end{equation}\normalsize where $\mathbf{a}$ is a vector of the pre-factors of the renormalization coefficients. 

To determine the value of $\tau$, a search procedure is performed over the range [-1,1] with an increment of 0.01 between successive values. For each value of $\tau$, the pre-factors are fit by minimizing Eq. (\ref{scaling_cost}) and then tabulated along with the corresponding error and $\tau$ value. The $\tau$ value and corresponding coefficients resulting in the minimum error are then extracted for evolving the ROM.

The estimation of the pre-factors is rather delicate. This is due to the rapid increase with $N$ of the condition number of the matrix of the least-squares problem \eqref{scaling_cost}. Our calculations are performed in double precision and this forces us to limit the estimation of the renormalization coefficients to only reduced models of size up to $N=14.$ 

When $N$ is small and correspondingly $M=2N$ is small, the estimation of $\tau$ is also delicate. The reason is that for small $M,$ the full order model cannot advance for long enough time so that a robust transfer of energy from the resolved to the unresolved variables can be established. As a result, we do not have enough accuracy to estimate $\tau$ reliably and its value fluctuates wildly with $M.$ That is why, while we have chosen to keep $M=2N,$ we use a simulation of size $M' > M$ to extract the quantities $\Delta E_k(t)$ needed for the estimation of the pre-factors in \eqref{scaling_cost}. We found that the optimal value of $\tau$ approaches an asymptotic value for $M'$ large (see Fig. C.5 for more details). In addition, we find that the asymptotic value of $\tau,$ depends weakly on $N$ (see Fig. C.6). Nevertheless, if we fix $\tau$ to the same value for all the different $N$ and plot the pre-factors $a_i$ as a function of $N$ we find robust scaling laws (see Table \ref{tab:Burgers_scaling_law_table} below where we have fixed $\tau=0.4$ for all $N$). These results were obtained for $M'=16384.$

\begin{table}[ht]
        \begin{center}
        \begin{adjustbox}{max width=\textwidth}
        \begin{tabular}{|c||c|c|c|c||c|c|c|c|}
        \hline$n$ & $\beta_1^n$& $\beta_2^n$& $\beta_3^n$& $\beta_4^n$& $\gamma_1^n$& $\gamma_2^n$& $\gamma_3^n$& $\gamma_4^n$\\
        \hhline{|=#=|=|=|=#=|=|=|=|}
        1 & $1.69$ &  &  &  & $-1.08$ &  &  &   \\
        \hline
        2 & $2.59$ & $-2.97$ &  &  & $-0.86$ & $-1.98$ &  &    \\
        \hline
        3 & $4.61$ & $-6.83$ & $4.15$ &  & $-0.97$ & $-1.96$ & $-3.04$ &    \\
        \hline
        4 & $6.16$ & $-10.89$ & $9.54$ & $-2.88$ & $-1.00$ & $-1.99$ & $-3.11$ & $-4.41$ \\
        \hline
        \end{tabular} 
        \end{adjustbox}
        \caption{Scaling laws for the pre-factors corresponding to  $\tau=0.4.$ The pre-factors are approximated by $a_i = \beta_i^nN^{\gamma_i^n}.$ We minimize \eqref{scaling_cost} with $n = 1,2,3,4$ and $N = 6,8,\dots,14$, then conduct a linear least squares fit of $\log(a_i)$ against $\log(N)$. The correlation coefficients for n = 1, 3, \& 4 fits were between 0.998 and 1 and the fits for n = 2 were between 0.941 and 0.972.} 
        \label{tab:Burgers_scaling_law_table} 
        \end{center}
    \end{table}

The results in Table \ref{tab:Burgers_scaling_law_table} lead to the following observations. First, we see that the exponents $\gamma_i^n$ seem relatively independent of the number of memory terms $n$ included in the reduced model. Thus, each additional memory term is making \emph{corrections} to previously captured behavior, but their contributions seem to be orthogonal to one another. Second, as seen by the negative exponents $\gamma_i^n,$ the magnitude of the pre-factors decreases with $N$  but the magnitude of the exponents $\gamma_i^n$ increases with $n.$ Taken together, these observations mean our renormalized expansion is indeed a perturbative one . We also see that the coefficients of the even terms are negative while the coefficients of the odd terms are positive in all cases. Thus, the renormalized coefficients agree with the unrenormalized coefficients $\frac{(-1)^{i+1}}{i!}$ in sign though not in magnitude.

\begin{figure}[h]
\begin{center}
 \includegraphics[width=.6\textwidth]{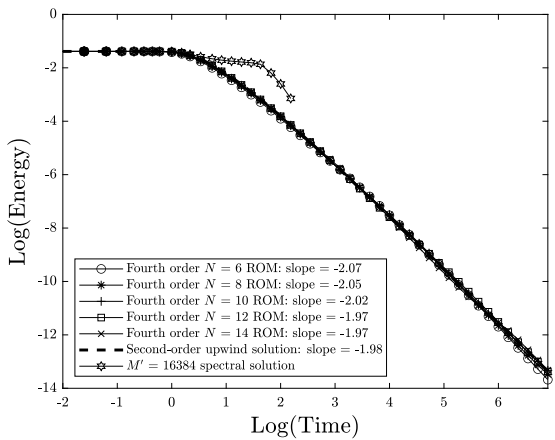}
\end{center}
\caption{The energy contained in the resolved modes of order $n=4$ ROMs described in \eqref{renormalized} and utilizing the optimal $\tau$ value for up to time $t=1000$ depicted on a log-log plot.} \label{fig:Burgers_long_energy}
\end{figure}

Fig. \ref{fig:Burgers_long_energy} shows a log-log plot of the evolution of the energy $\frac{1}{2}\sum_{k \in F} E_k(t)$ contained in the resolved modes of of order $n=4$ ROMs of different size $N.$ For each of these ROMs, the optimal value of $\tau$ has been used. Also, due to the perturbative structure of the ROM, the $n=4$ ROM results are converged (see Figs. C.1 and C.2 for the evolution of the energy and relative error for fixed size $N=14,$ optimal $\tau$ and increased ROM order).

The predictions of the ROMs are compared to that of a second-order upwind scheme with $\Delta x = \frac{2 \pi}{10000}.$ It is remarkable that the ROMs, whose coefficients were calibrated using only full order model data \emph{before} the occurrence of the shock at $T=1,$ can actually predict with high accuracy the evolution of the energy for such long times (see also Fig. C.3 for the evolution of the relative error in the prediction of the energy). The slope was calculated using data in the time window $15 \leq t \leq 500.$ The exact slope value is -2 (see e.g. \cite{lax1973hyperbolic}).

\begin{figure}[h]
\begin{center}
 \includegraphics[width=0.6\textwidth]{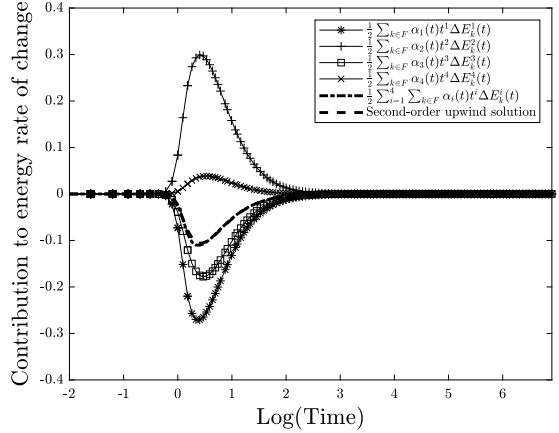}
\end{center}
\caption{The contribution of each order memory term to the rate of change of the energy in the resolved modes for the N = 14 fourth order ROM utilizing the optimal $\tau$ value for up to time $t=1000$.}\label{fig:Burgers_long_memory}
\end{figure}

Fig. \ref{fig:Burgers_long_memory} shows the contribution $\alpha_i(t) t^i \Delta E_{k}^i(t),$ $i=1,\ldots,4$ of the memory terms to the rate of change of the energy in the resolved modes for the N = 14 fourth order ROM with optimal $\tau \approx 0.33.$ We see that the contributions of the first and second order terms are comparable, while those of the third and fourth order terms are significantly smaller.  Also, the first and third order contributions are negative definite, while the second and fourth are positive definite (see also Fig. C.4 for the prediction of the real space solution for different instants).

\section{Renormalized ROMs for the 3D Euler equations}\label{euler}

The Euler equations are given by:
\begin{equation}
\mathbf{u}_t + \mathbf{u}\cdot\nabla\mathbf{u} = -\nabla p,\quad \nabla\cdot \mathbf{u} = 0\label{Euler}
\end{equation} where $\mathbf{u}(\mathbf{x},t)$ is the three-dimensional velocity field and $p$ is the pressure. We consider the solution in a periodic box $[0,2\pi]^3$ with the Taylor-Green initial condition, $\mathbf{u}=(\sin(x_1)\cos(x_2)\cos(x_3),-\cos(x_1)\sin(x_2)\cos(x_3),0)^T.$ We write $\mathbf{u}$ as a Fourier series:
$
\mathbf{u}(x,t) = \sum_{\mathbf{k}} \mathbf{u}_{\mathbf{k}}(t) e^{i\mathbf{k}\cdot\mathbf{x}}\label{eq:fourier}
$
where $\mathbf{k}$ is a three-dimensional wavevector, and the sum is over all possible integer-valued wavevectors. The evolution of a Fourier mode is:
\begin{equation}
\frac{d\mathbf{u}_{\mathbf{k}}}{dt} =\mathbf{R}_\mathbf{k}(\mathbf{u}) =  -i\sum_{\mathbf{p}+\mathbf{q}=\mathbf{k}}\mathbf{k}\cdot\mathbf{u}_{\mathbf{p}}A_{\mathbf{k}}\mathbf{u}_{\mathbf{q}}\label{evolution}
\end{equation}where $A_{\mathbf{k}} = I - \frac{\mathbf{k}\mathbf{k}^T}{|\mathbf{k}|^2}$ is the incompressibility projection operator \cite{doering1995applied}.  Consider the Fourier components $\mathbf{u}_{\mathbf{k}}(t)$, where $\mathbf{k}\in F\cup G$. Let $F$ be the set of resolved modes. That is $F = \{\mathbf{k} \in [-N+1,N-1]^3\}$. Let  $F\cup G = \{\mathbf{k} \in [-M+1,M-1]^3\}$. Define $\hat{\mathbf{u}} = \{\mathbf{u}_{\mathbf{k}}\;\; |\;\; \mathbf{k}\in F\}$ and $\tilde{\mathbf{u}} = \{\mathbf{u}_{\mathbf{k}}\;\; | \;\;\mathbf{k} \in G\}$. We define $\mathcal{L}$ such that $\frac{d\mathbf{u}_{\mathbf{k}}}{dt} = e^{t\mathcal{L}}\mathcal{L}\mathbf{u}_{\mathbf{k}}^0$ and the projection operator $P$ such that $P f(\mathbf{u}) = P f(\hat{\mathbf{u}}^0,\tilde{\mathbf{u}}^0) = f(\hat{\mathbf{u}}^0,0)$ once again. With these definitions, we can construct the terms of the CMA $\mathbf{R}_\mathbf{k}^i(\hat{\mathbf{u}})$.

With the exception of the $t$-model, the resulting unrenormalized reduced order models are not stable, so we again renormalize by fitting renormalization coefficients of the form $\alpha_i(t) = a_it^{-i\tau}$ by minimizing an error term equivalent to Eq. (\ref{scaling_cost}) (but with $\mathbf{k}$ as a wavevector instead of wavenumber). Again, as in the Burgers case, we will renormalize against data $\Delta E_{\mathbf{k}}(t)$ produced by a full model that we trust has not yet become unresolved.

We ran a full simulation of size $M'=48$ in each of the three directions and used it to compute renormalization coefficients for reduced order models of size $N = 6,\dots,14.$ For each value of $N$, we considered ROMs that included up through $n = 1,\ldots,4$ terms from the CMA. The restriction of the size $N$ to only up to 14 was dictated again by the high condition number of the matrix in the least-squares problem. Also, we restricted the size of $M'$ to small values due to a limited computational capacity. To utilize considerably larger values of $M'$ will require parallelization of the code and more powerful computational resources.  

Due to the use of a small value for $M',$ we do not get a reliable estimate for the optimal value of  $\tau$ if we follow the same procedure as for Burgers. Also, we found that unlike Burgers, approximately for $\tau \in [0,.4],$ the renormalized ROMs are unstable. This means that the renormalization of 3D Euler is more nuanced than Burgers. This is most likely due to the formation of small scale structures which are more complex than a shock. Since we cannot presently decide on an optimal value of $\tau,$ we show results only for $\tau=1$ (see Fig. C.10 for results with different values of $\tau.$)   

As in the case of Burgers, if we fix the value of $\tau,$ the pre-factors of the renormalized coefficients do follow robust scaling laws as a function of the size $N$ (see Fig. C.11 and Table \ref{tab:scaling_law_table} for $\tau=1$).
\begin{table}[ht]
    \begin{center}
    \begin{adjustbox}{max width=\textwidth}
    \begin{tabular}{|c||c|c|c|c||c|c|c|c|}
    \hline$n$ & $\beta_1^n$& $\beta_2^n$& $\beta_3^n$& $\beta_4^n$& $\gamma_1^n$& $\gamma_2^n$& $\gamma_3^n$& $\gamma_4^n$\\
    \hhline{|=#=|=|=|=#=|=|=|=|}
    1 & $1.62$ &  &  &  & $-1.09$ &  &  &   \\
    \hline
    2 & $2.43$ & $-2.32$ &  &  & $-1.01$ & $-2.17$ &  &    \\
    \hline
    3 & $2.76$ & $-3.47$ & $1.65$ &  & $-0.98$ & $-2.11$ & $-3.37$ &    \\
    \hline
    4 & $2.95$ & $-4.46$ & $4.16$ & $-1.53$ & $-0.90$ & $-1.91$ & $-3.06$ & $-4.27$ \\
    \hline
    \end{tabular} 
    \end{adjustbox}
    \caption{Scaling laws that approximate the pre-factors corresponding to $\tau$ = 1. The pre-factors are approximated by $a_i = \beta_i^nN^{\gamma_i^n}.$ We minimize \eqref{scaling_cost} with $n = 1,2,3,4$ and $N = 6,8,\dots,14$, then conduct a linear least squares fit of $\log(a_i)$ against $\log(N)$.The correlation coefficient for all fits was between 0.998 and 1.} \label{tab:scaling_law_table} 
    \end{center}
    \end{table}

The behavior of the solution to the 3D Euler equations with a smooth initial condition like Taylor-Green remains unknown. Consequently, we cannot compare the results of our ROMs to the exact solution for validation. Instead, we endeavour to produce ROMs that remain \emph{stable} over a long time. We will have to rely upon secondary means of inferring the accuracy of the resultant ROMs. Our results, not fully validated as they are, can be interpreted as evidence that is suggestive of long-term behavior of a subset of Fourier modes evolved according to Euler's equations.

Our renormalized ROMs led to solutions that remained stable until at least $t=1000$. When we fix a resolution $N$ and simulate ROMs that include up through order $n=1,2,3,4$, the results appear to converge with increasing order (see Fig. C.9 for more details). This strengthens our assessment of the perturbative nature of our expansion. Each additional term in a ROM is more expensive to compute, and the fast convergence gives us confidence that including additional terms will only minimally affect our results. Thus, we will assume that the fourth order ROMs represent the most accurate simulations of the dynamics of the resolved modes. 

\begin{figure}[h]
\begin{center}
 \includegraphics[width=0.6\textwidth]{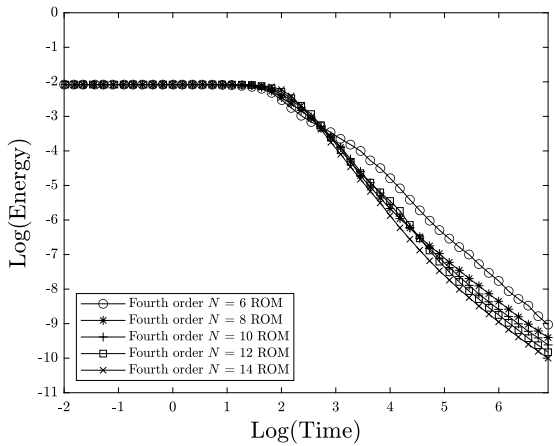}
\end{center}
\caption{The energy contained in the resolved modes of fourth order ROMs up to time $t=1000$ depicted on a log-log plot.}\label{fig:long_energy}
\end{figure}

Fig. \ref{fig:long_energy} depicts the energy decay of fourth order ROMs with resolution $N=6,\ldots,14$ up to time $t=1000$ on a log-log plot. We see that in all cases there is monotonic energy decay. As time goes on, the results become stratified: the amount of energy remaining in the system \emph{decreases} with \emph{increasing} ROM resolution. This indicates significant activity in the high-frequency modes that increases with the resolution.

The decay of energy indicates the presence of two different regimes of algebraic (in time) energy ejection from the resolved modes (we note that the existence of two different energy decay regimes has been put forth in \cite{speziale}). Table \ref{tab:energy_decay_rates} shows the slope from the data for which between 50\% and 90\% of the initial energy has left the system (the initial decay rate). We see that the rate of energy ejection eventually becomes slightly smaller. We computed the slope from the data after 99.5\% of the initial energy had left the system (the second decay rate). We have also included in Table \ref{tab:energy_decay_rates} an estimate of the initial decay time, which we have defined as the time that 10\% of the energy has left the resolved modes.

\begin{table}[ht]
        \begin{center}
        \begin{adjustbox}{max width=\textwidth}
        \begin{tabular}{|c||c|c|c|}
        \hline
        ROM Resolution N & Initial Decay Time & Initial Decay Rate & Second Decay Rate \\
        \hhline{|=#=|=|=|}
        $6$ & $4.946$ & $-0.998$ & $-1.353$ \\
        \hline
        $8$ & $5.411$ & $-1.580$ & $-1.195$ \\
        \hline
        $10$ & $5.810$ & $-1.649$ & $-1.215$ \\
        \hline
        $12$ & $6.138$ & $-1.907$ & $-1.264$ \\
        \hline
        $14$ & $6.853$ & $-1.980$ & $-1.258$ \\
        \hline
        \end{tabular} 
        \end{adjustbox}
        \caption{Energy decay rates of fourth order ROMs using the renormalization coefficients as described in Table \ref{tab:scaling_law_table} (see text for details).} 
        \label{tab:energy_decay_rates} 
        \end{center}
    \end{table}

\begin{figure}[h]
\begin{center}
 \includegraphics[width=0.6\textwidth]{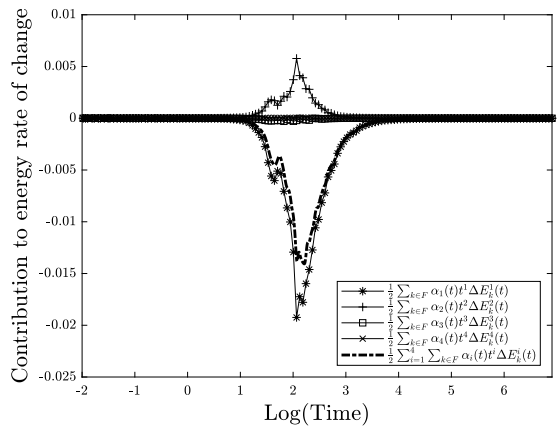}
\end{center}
\caption{The contribution of each order memory term to the rate of change of the energy in the resolved modes for the N = 14 fourth order ROM with $\tau=1.0.$}\label{fig:Euler_long_memory}
\end{figure}

Fig. \ref{fig:Euler_long_memory} presents the evolution of the contribution of the various memory terms to the rate of change of the energy in the resolved modes. The perturbative nature of our approach is evident in the stratification of the contributions of the various memory terms (see also Figs. C.12-C.13 and Table C.1). We note that our results indicate a peak for the rate of energy rate of change around time $t=8-9$ which is in agreement with recent very large scale direct simulations (see Fig. 8 in \cite{fehn2020numerical}).

\section{Discussion}

We have presented a novel way of controlling the memory length of renormalized ROMs for multi-scale systems whose brute force simulation can be prohibitively expensive. We have validated our approach for the inviscid Burgers equation, where our perturbatively renormalized ROMs can make predictions of remarkable accuracy for long times. 

Furthermore, we have presented results for the 3D Euler equations of incompressible fluid flow, where we have obtained stable results for long times. Despite the wealth of theoretical and numerical studies, the exact behavior of solutions to the 3D Euler equations is unknown (see a very partial list  \cite{marchioro1994euler,Shnirelman2000,majda,constantin2017analysis,isett2017onsager,elgindi2019finite,grafke2013lagrangian,luo2014potentially,agafontsev2015development,Larios2018}). Even modern simulations with exceptionally high resolution cannot proceed for long times. Thus, our ROMs represent an advancement in the ability to simulate these equations. Without an exact solution to validate against, it is difficult to ascertain whether our results are accurate in addition to stable. However, there are a few hints: the convergence of behavior with increasing order indicates that our ROMs have a perturbative structure. That is, each additional order in the ROM modifies the solution less and less. Next, Table \ref{tab:scaling_law_table} demonstrates that adding additional terms does not significantly change the scaling laws for the previous terms. Each additional term is making corrections to previously captured behavior. These observations together give us reason to trust these results.

The perturbative renormalization of our ROMs is possible due to the smoothness of the used initial condition. By smoothness we mean the ratio of the highest wavenumber active in the initial condition (1 in our examples), over the highest wavenumber that can be resolved by the ROM (N in our examples). This ratio, $1/N$ in our case, serves as the small quantity for the perturbation expansion. The reason can be found in the structure of the expressions for the memory terms as we increase the order. These expressions, when transformed back to physical space, involve higher and higher order derivatives. Thus, they probe smaller and smaller scales. For a smooth initial condition (small ratio), such expressions involving higher derivatives need only to contribute a little to capture the transfer of energy out of the resolved modes. As a result, they acquire \emph{renormalized} coefficients of \emph{decreasing} magnitude as we go up in order. This observation creates an interesting analogy between our approach and perturbatively renormalizable diagrammatic expansions in high energy physics, as well as the perturbative renormalization of computations based on Kolmogorov complexity \cite{manin2014complexity}. In essence, we can think of CMA as an expansion of the memory in terms of increasing Kolmogorov complexity (see expressions in Appendix), whose importance in the representation of the memory for a smooth initial condition decreases with order.

We plan to apply our framework to larger simulations of 3D Euler (larger $M'$) so that we can determine an optimal value of $\tau.$ In addition, to utilize the extracted scaling laws to simulate ROMs for higher resolutions where the direct estimation of the renormalization coefficients may not be possible (see Figs. C.7-C.8 for preliminary results for Burgers and Fig. C.14 for 3D Euler). We have also obtained results for the 2D Euler equations which have a very different behavior and we will present those elsewhere. Finally, it will be very interesting to investigate how adding viscous dissipation (Navier-Stokes equations) will alter the scaling dependence of renormalized coefficients, including the likely occurrence of incomplete similarity \cite{barenblatt,chorin2005spectra,price2019renormalized}.

\section{Acknowledgements}

The work of PS was supported by the U.S. Department of Energy (DOE) Office of Science, Office of Advanced Scientific Computing Research (ASCR) as part of the Multifaceted Mathematics for Rare, Extreme Events in Complex Energy and Environment Systems (MACSER) project. Pacific Northwest National Laboratory is operated by Battelle for the DOE under Contract DE-AC05-76RL01830.

\bibliographystyle{plain}
\bibliography{MultiScale}

\newpage

\appendix 

\renewcommand\thefigure{\thesection.\arabic{figure}}    
\setcounter{figure}{0}  
\renewcommand\thetable{\thesection.\arabic{table}}    
\setcounter{table}{0}  

\section{The complete memory approximation of the 3D Euler equations}
In the main text, we derived rules for applying the Liouvillian $\mathcal{L}$ and the projector $P$ to expressions. We can use these rules to derive the second, third, and fourth order terms of the complete memory approximation of the 3D Euler equations. The expressions for the  memory terms for Burgers can be derived in a similar fashion. 

We define a convolution operator
\begin{equation}\mathbf{C}_{\mathbf{k}}(\mathbf{v},\mathbf{w}) = -i\sum_{\substack{\mathbf{p}+\mathbf{q}=\mathbf{k}\\\mathbf{p},\mathbf{q} \in F\cup G}}\mathbf{k}\cdot\mathbf{v}_{\mathbf{p}} A_{\mathbf{k}} \mathbf{w}_{\mathbf{q}}.
\end{equation}We will now also define $\mathbf{D}_\mathbf{k}(\mathbf{v},\mathbf{w}) = \mathbf{C}_{\mathbf{k}}(\mathbf{v},\mathbf{w})+\mathbf{C}_{\mathbf{k}}(\mathbf{w},\mathbf{v})$ and the related convolutions $\hat{\mathbf{D}}$ and $\tilde{\mathbf{D}}$ defined as the resolved and unresolved modes of $\mathbf{D}_\mathbf{k}$, respectively. These will be useful in simplifying notation.

The second order term of the complete memory approximation is:
\begin{equation}\mathbf{R}^2_{\mathbf{k}}(\hat{\mathbf{u}}) = Pe^{t\mathcal{L}}P\mathcal{L}[P\mathcal{L}-Q\mathcal{L}]Q\mathcal{L}\mathbf{u}_{\mathbf{k}}^0.
\end{equation}First note that we can use the symmetric convolution function $\mathbf{D}$ to write \begin{equation}Q\mathcal{L}\mathbf{u}_{\mathbf{k}}^0 = \mathbf{D}_{\mathbf{k}}(\hat{\mathbf{u}},\tilde{\mathbf{u}}) + \mathbf{C}_{\mathbf{k}}(\tilde{\mathbf{u}},\tilde{\mathbf{u}}).\end{equation} $\mathcal{L}$ operates upon $\mathbf{D}$ in the same manner it did upon $\mathbf{C}$. The projector $P$ when applied to $\mathbf{D}$ similarly is applied to each term within the expression. Starting from this, we derive an expression for the $t^2$-term:
\begin{align}
&\mathbf{R}^2_{\mathbf{k}}(\hat{\mathbf{u}}) = Pe^{t\mathcal{L}}P\mathcal{L}(2P\mathcal{L}-\mathcal{L})[\mathbf{D}_{\mathbf{k}}(\hat{\mathbf{u}}^0,\tilde{\mathbf{u}}^0)+\mathbf{C}_{\mathbf{k}}(\tilde{\mathbf{u}}^0,\tilde{\mathbf{u}}^0)]\notag\\
&=Pe^{t\mathcal{L}}P\mathcal{L}[2\mathbf{D}_{\mathbf{k}}(\hat{\mathbf{u}}^0,\tilde{\mathbf{C}}(\hat{\mathbf{u}}^0,\hat{\mathbf{u}}^0))-\mathbf{D}_{\mathbf{k}}(\hat{\mathbf{C}}(\mathbf{u}^0,\mathbf{u}^0),\tilde{\mathbf{u}}^0)\notag\\
&\qquad\qquad\quad -\mathbf{D}_{\mathbf{k}}(\hat{\mathbf{u}}^0,\tilde{\mathbf{C}}(\mathbf{u}^0,\mathbf{u}^0))-\mathbf{D}_{\mathbf{k}}(\tilde{\mathbf{u}}^0,\tilde{\mathbf{C}}(\mathbf{u}^0,\mathbf{u}^0))]\notag\\
&=Pe^{t\mathcal{L}}[\mathbf{D}_{\mathbf{k}}(\hat{\mathbf{u}}^0,\tilde{\mathbf{D}}(\hat{\mathbf{C}}(\hat{\mathbf{u}}^0,\hat{\mathbf{u}}^0)-\tilde{\mathbf{C}}(\hat{\mathbf{u}}^0,\hat{\mathbf{u}}^0),\hat{\mathbf{u}}^0))\notag\\
&\qquad \qquad -\mathbf{D}_{\mathbf{k}}(\tilde{\mathbf{C}}(\hat{\mathbf{u}}^0,\hat{\mathbf{u}}^0),\tilde{\mathbf{C}}(\hat{\mathbf{u}}^0,\hat{\mathbf{u}}^0))]\notag\\
&=\mathbf{D}_{\mathbf{k}}(\hat{\mathbf{u}},\tilde{\mathbf{D}}(\hat{\mathbf{C}}(\hat{\mathbf{u}},\hat{\mathbf{u}})-\tilde{\mathbf{C}}(\hat{\mathbf{u}},\hat{\mathbf{u}}),\hat{\mathbf{u}})) \notag \\ & \quad -\mathbf{D}_{\mathbf{k}}(\tilde{\mathbf{C}}(\hat{\mathbf{u}},\hat{\mathbf{u}}),\tilde{\mathbf{C}}(\hat{\mathbf{u}},\hat{\mathbf{u}})).
\end{align}

We will make use of the terms we have already computed in order to simplify our derivation of the third order term. Under the complete memory approximation, the third term is:
\begin{gather}
\mathbf{R}^3_{\mathbf{k}}(\hat{\mathbf{u}})  =  Pe^{t\mathcal{L}}P\mathcal{L}[PLPL-2PLQL \notag \\
-2QLPL+QLQL]Q\mathcal{L}\mathbf{u}_{\mathbf{k}}^0.
\end{gather}
We rewrite it with no $Q\mathcal{L}$ terms.
\begin{align*}
\mathbf{R}^3_{\mathbf{k}}(\hat{\mathbf{u}})  =&  Pe^{t\mathcal{L}}P\mathcal{L}[PLPL-2PLQL-2QLPL \\
& 
+QLQL]Q\mathcal{L}\mathbf{u}_{\mathbf{k}}^0\\
=&Pe^{t\mathcal{L}}P\mathcal{L}[3P\mathcal{L}(2P\mathcal{L}-P\mathcal{L})-3\mathcal{L}P\mathcal{L}+\mathcal{L}\mathcal{L}]Q\mathcal{L}\mathbf{u}_{\mathbf{k}}^0
\end{align*}We recognize that we have already computed an expression for $P\mathcal{L}(2P\mathcal{L}-\mathcal{L})Q\mathcal{L}\mathbf{u}_{\mathbf{k}}^0$ during our derivation of the second order term. This leaves two additional terms to compute before simplifying and applying the final $P\mathcal{L}$:
\begin{align*}
\mathcal{L}P\mathcal{L}Q\mathcal{L}\mathbf{u}_{\mathbf{k}}^0 =&  \mathcal{L}[\mathbf{D}_{\mathbf{k}}(\hat{\mathbf{u}}^0,\tilde{\mathbf{C}}(\hat{\mathbf{u}}^0,\hat{\mathbf{u}}^0))]\\
=&\mathbf{D}_{\mathbf{k}}(\hat{\mathbf{C}}(\mathbf{u}^0,\mathbf{u}^0),\tilde{\mathbf{C}}(\hat{\mathbf{u}}^0,\hat{\mathbf{u}}^0))\\
& 
+\mathbf{D}_{\mathbf{k}}(\hat{\mathbf{u}}^0,\tilde{\mathbf{D}}(\hat{\mathbf{C}}(\mathbf{u}^0,\mathbf{u}^0),\hat{\mathbf{u}}^0))
\end{align*}
and
\begin{align*}
\mathcal{L}\mathcal{L}Q\mathcal{L}\mathbf{u}_{\mathbf{k}}^0 = &\mathcal{L}\mathcal{L}[\mathbf{D}_{\mathbf{k}}(\hat{\mathbf{u}}^0,\tilde{\mathbf{u}}^0)+\mathbf{C}_{\mathbf{k}}(\tilde{\mathbf{u}}^0,\tilde{\mathbf{u}}^0)]\\
=&\mathcal{L}[\mathbf{D}_{\mathbf{k}}(\hat{\mathbf{C}}(\mathbf{u}^0,\mathbf{u}^0),\tilde{\mathbf{u}}^0)+\mathbf{D}_{\mathbf{k}}(\hat{\mathbf{u}}^0,\tilde{\mathbf{C}}(\mathbf{u}^0,\mathbf{u}^0))\\
&  
+\mathbf{D}_{\mathbf{k}}(\tilde{\mathbf{C}}(\mathbf{u}^0,\mathbf{u}^0),\tilde{\mathbf{u}}^0)]\\
=& \mathbf{D}_{\mathbf{k}}(\hat{\mathbf{D}}(\mathbf{C}(\mathbf{u}^0,\mathbf{u}^0),\mathbf{u}^0),\tilde{\mathbf{u}}^0)\\
&  
+2\mathbf{D}_{\mathbf{k}}(\hat{\mathbf{C}}(\mathbf{u}^0,\mathbf{u}^0),\tilde{\mathbf{C}}(\mathbf{u}^0,\mathbf{u}^0))\\
&+ \mathbf{D}_{\mathbf{k}}(\mathbf{u}^0,\tilde{\mathbf{D}}(\mathbf{C}(\mathbf{u}^0,\mathbf{u}^0),\mathbf{u}^0))\\
&  
+\mathbf{D}_{\mathbf{k}}(\tilde{\mathbf{C}}(\mathbf{u}^0,\mathbf{u}^0),\tilde{\mathbf{C}}(\mathbf{u}^0,\mathbf{u}^0)).
\end{align*}Combining these three computed terms with the correct coefficients yields:
\begin{gather}
[3P\mathcal{L}(2P\mathcal{L}-P\mathcal{L})-3\mathcal{L}P\mathcal{L}+\mathcal{L}\mathcal{L}]Q\mathcal{L}\mathbf{u}_{\mathbf{k}}^0= \notag \\
 \mathbf{D}_{\mathbf{k}}(\hat{\mathbf{u}}^0,\tilde{\mathbf{D}}(-3\hat{\mathbf{C}}(\mathbf{u}^0,\mathbf{u}^0)+3\hat{\mathbf{C}}(\hat{\mathbf{u}}^0,\hat{\mathbf{u}}^0)-3\tilde{\mathbf{C}}(\hat{\mathbf{u}}^0,\hat{\mathbf{u}}^0),\hat{\mathbf{u}}^0)) \notag \\
-6\mathbf{C}_{\mathbf{k}}(\tilde{\mathbf{C}}(\hat{\mathbf{u}}^0,\hat{\mathbf{u}}^0),\tilde{\mathbf{C}}(\hat{\mathbf{u}}^0,\hat{\mathbf{u}}^0))-3\mathbf{D}_{\mathbf{k}}(\hat{\mathbf{C}}(\mathbf{u}^0,\mathbf{u}^0),\tilde{\mathbf{C}}(\hat{\mathbf{u}}^0,\hat{\mathbf{u}}^0)) \notag \\
+\mathbf{D}_{\mathbf{k}}(\hat{\mathbf{D}}(\mathbf{C}(\mathbf{u}^0,\mathbf{u}^0),\mathbf{u}^0),\tilde{\mathbf{u}}^0)+2\mathbf{D}_{\mathbf{k}}(\hat{\mathbf{C}}(\mathbf{u}^0,\mathbf{u}^0),\tilde{\mathbf{C}}(\mathbf{u}^0,\mathbf{u}^0)) \notag \\
+ \mathbf{D}_{\mathbf{k}}(\mathbf{u}^0,\tilde{\mathbf{D}}(\mathbf{C}(\mathbf{u}^0,\mathbf{u}^0),\mathbf{u}^0))+2\mathbf{C}_{\mathbf{k}}(\tilde{\mathbf{C}}(\mathbf{u}^0,\mathbf{u}^0),\tilde{\mathbf{C}}(\mathbf{u}^0,\mathbf{u}^0)).
\end{gather}

The $t^3$-term is found once we apply $P\mathcal{L}$ to this expression, yielding:
\begin{align*}
\mathbf{R}_{\mathbf{k}}^3(\hat{\mathbf{u}}) 
=&\mathbf{D}_{\mathbf{k}}(\hat{\mathbf{u}},\tilde{\mathbf{D}}(\hat{\mathbf{u}},\hat{\mathbf{D}}(\hat{\mathbf{u}},\hat{\mathbf{C}}(\hat{\mathbf{u}},\hat{\mathbf{u}})-2\tilde{\mathbf{C}}(\hat{\mathbf{u}},\hat{\mathbf{u}}))\\
& +\tilde{\mathbf{D}}(\hat{\mathbf{u}},\tilde{\mathbf{C}}(\hat{\mathbf{u}},\hat{\mathbf{u}})-2\hat{\mathbf{C}}(\hat{\mathbf{u}},\hat{\mathbf{u}})))\\
&+\tilde{\mathbf{D}}(\tilde{\mathbf{C}}(\hat{\mathbf{u}},\hat{\mathbf{u}}),\tilde{\mathbf{C}}(\hat{\mathbf{u}},\hat{\mathbf{u}})-\hat{\mathbf{C}}(\hat{\mathbf{u}},\hat{\mathbf{u}}))\\
& +\tilde{\mathbf{D}}(\hat{\mathbf{C}}(\hat{\mathbf{u}},\hat{\mathbf{u}}),\hat{\mathbf{C}}(\hat{\mathbf{u}},\hat{\mathbf{u}})))\\
&+3\mathbf{D}_{\mathbf{k}}(\tilde{\mathbf{C}}(\hat{\mathbf{u}},\hat{\mathbf{u}}),\tilde{\mathbf{D}}(\hat{\mathbf{u}},\tilde{\mathbf{C}}(\hat{\mathbf{u}},\hat{\mathbf{u}})-\hat{\mathbf{C}}(\hat{\mathbf{u}},\hat{\mathbf{u}}))).
\end{align*}

We will include up through the fourth-order term in our renormalized ROMs. The derivation of these models is quite tedious. For this reason, we make use of our symbolic tools in \cite{MZrepos}. The result for the fourth order model is:

\begin{align*}
\mathbf{R}_{\mathbf{k}}^4(\hat{\mathbf{u}})=&e^{t\mathcal{L}}P\mathcal{L}[PLPLPL - 3 PLPLQL - 
 5 PLQLPL \\
 & +3 PLQLQL  -3 QLPLPL + 5 QLPLQL \\
 & +3 QLQLPL - QLQLQL]Q\mathcal{L}\mathbf{u}_{\mathbf{k}}^0\\
 =&
\mathbf{D}_{\mathbf{k}}(\hat{\mathbf{u}},\tilde{\mathbf{D}}(\hat{\mathbf{u}},\hat{\mathbf{D}}(\hat{\mathbf{C}}(\hat{\mathbf{u}},\hat{\mathbf{u}}),\hat{\mathbf{C}}(\hat{\mathbf{u}},\hat{\mathbf{u}})-2\tilde{\mathbf{C}}(\hat{\mathbf{u}},\hat{\mathbf{u}}))\\
& +3\hat{\mathbf{D}}(\tilde{\mathbf{C}}(\hat{\mathbf{u}},\hat{\mathbf{u}}),\tilde{\mathbf{C}}(\hat{\mathbf{u}},\hat{\mathbf{u}}))\\
&+\tilde{\mathbf{D}}(\hat{\mathbf{C}}(\hat{\mathbf{u}},\hat{\mathbf{u}}),2\tilde{\mathbf{C}}(\hat{\mathbf{u}},\hat{\mathbf{u}})-3\hat{\mathbf{C}}(\hat{\mathbf{u}},\hat{\mathbf{u}}))\\
&-\tilde{\mathbf{D}}
(\tilde{\mathbf{C}}(\hat{\mathbf{u}},\hat{\mathbf{u}}),\tilde{\mathbf{C}}(\hat{\mathbf{u}},\hat{\mathbf{u}}))\\
&+\hat{\mathbf{D}}(\hat{\mathbf{u}},\hat{\mathbf{D}}(\hat{\mathbf{u}},\hat{\mathbf{C}}(\hat{\mathbf{u}},\hat{\mathbf{u}})-3\tilde{\mathbf{C}}(\hat{\mathbf{u}},\hat{\mathbf{u}}))\\
& +\tilde{\mathbf{D}}(\hat{\mathbf{u}},3\tilde{\mathbf{C}}(\hat{\mathbf{u}},\hat{\mathbf{u}})-5\hat{\mathbf{C}}(\hat{\mathbf{u}},\hat{\mathbf{u}})))\\
&+\tilde{\mathbf{D}}(\hat{\mathbf{u}},\hat{\mathbf{D}}(\hat{\mathbf{u}},5\tilde{\mathbf{C}}(\hat{\mathbf{u}},\hat{\mathbf{u}})-3\hat{\mathbf{C}}(\hat{\mathbf{u}},\hat{\mathbf{u}}))\\
& +\tilde{\mathbf{D}}(\hat{\mathbf{u}},3\hat{\mathbf{C}}(\hat{\mathbf{u}},\hat{\mathbf{u}})-\tilde{\mathbf{C}}(\hat{\mathbf{u}},\hat{\mathbf{u}}))))\\
& +\tilde{\mathbf{D}}(\hat{\mathbf{C}}(\hat{\mathbf{u}},\hat{\mathbf{u}}),\hat{\mathbf{D}}(\hat{\mathbf{u}},3\hat{\mathbf{C}}(\hat{\mathbf{u}},\hat{\mathbf{u}})-5\tilde{\mathbf{C}}(\hat{\mathbf{u}},\hat{\mathbf{u}}))\\
& +\tilde{\mathbf{D}}(\hat{\mathbf{u}},\tilde{\mathbf{C}}(\hat{\mathbf{u}},\hat{\mathbf{u}})-3\hat{\mathbf{C}}(\hat{\mathbf{u}},\hat{\mathbf{u}})))\\
& +\tilde{\mathbf{D}}(\tilde{\mathbf{C}}(\hat{\mathbf{u}},\hat{\mathbf{u}}),\hat{\mathbf{D}}(\hat{\mathbf{u}},3\tilde{\mathbf{C}}(\hat{\mathbf{u}},\hat{\mathbf{u}})-\hat{\mathbf{C}}(\hat{\mathbf{u}},\hat{\mathbf{u}}))\\
& +\tilde{\mathbf{D}}(\hat{\mathbf{u}},5\hat{\mathbf{C}}(\hat{\mathbf{u}},\hat{\mathbf{u}})-3\tilde{\mathbf{C}}(\hat{\mathbf{u}},\hat{\mathbf{u}}))))\\
&-4\mathbf{D}_{\mathbf{k}}(\tilde{\mathbf{C}}(\hat{\mathbf{u}},\hat{\mathbf{u}}),\tilde{\mathbf{D}}(\hat{\mathbf{C}}(\hat{\mathbf{u}},\hat{\mathbf{u}}),\hat{\mathbf{C}}(\hat{\mathbf{u}},\hat{\mathbf{u}})-\tilde{\mathbf{C}}(\hat{\mathbf{u}},\hat{\mathbf{u}}))\\
& +\tilde{\mathbf{D}}(\tilde{\mathbf{C}}(\hat{\mathbf{u}},\hat{\mathbf{u}}),\tilde{\mathbf{C}}(\hat{\mathbf{u}},\hat{\mathbf{u}}))\\
& +\tilde{\mathbf{D}}(\hat{\mathbf{u}},\hat{\mathbf{D}}(\hat{\mathbf{u}},\hat{\mathbf{C}}(\hat{\mathbf{u}},\hat{\mathbf{u}})-2\tilde{\mathbf{C}}(\hat{\mathbf{u}},\hat{\mathbf{u}}))\\
& +\tilde{\mathbf{D}}(\hat{\mathbf{u}},\tilde{\mathbf{C}}(\hat{\mathbf{u}},\hat{\mathbf{u}})-2\hat{\mathbf{C}}(\hat{\mathbf{u}},\hat{\mathbf{u}}))))\\
&-3\mathbf{D}_{\mathbf{k}}(\tilde{\mathbf{D}}(\hat{\mathbf{u}},\hat{\mathbf{C}}(\hat{\mathbf{u}},\hat{\mathbf{u}})),\tilde{\mathbf{D}}(\hat{\mathbf{u}},\hat{\mathbf{C}}(\hat{\mathbf{u}},\hat{\mathbf{u}})-2\tilde{\mathbf{C}}(\hat{\mathbf{u}},\hat{\mathbf{u}})))\\
&-3\mathbf{D}_{\mathbf{k}}(\tilde{\mathbf{D}}(\hat{\mathbf{u}},\tilde{\mathbf{C}}(\hat{\mathbf{u}},\hat{\mathbf{u}})),\tilde{\mathbf{D}}(\hat{\mathbf{u}},\tilde{\mathbf{C}}(\hat{\mathbf{u}},\hat{\mathbf{u}})))
.
\end{align*}

\section{Cost of simulating the ROMs for 3D Euler}
Each convolution requires a three-dimensional FFT and IFFT. Furthermore, for a system of resolution $N$, the FFTs are of size $(2\times 2\times 3/2 \times N)^3$. One factor of 2 comes from the fact that we include both positive and negative modes. The other factor of 2 because the convolutions require an intermediary ``full'' system double the size. The factor of $3/2$ is needed to dealias the results.  The Markov model requires only one FFT. The first order ROM requires three convolutions. The fourth order ROM requires 38 convolutions. Thus, the simulation cost grows very quickly as the order of the ROM increases. On the other hand, we are \emph{incapable} of performing a brute force calculation beyond a few units of time even on modern cutting-edge high-resolution simulations. Reduced order models allow us to utilize the multiscale structure of problems to evolve only a subset of the variables in the system. With our models we can integrate out to long times in only a few days on a laptop computer.

\section{Supplementary figures}

\begin{figure}[h]
\begin{center}
 \includegraphics[width=0.6\textwidth]{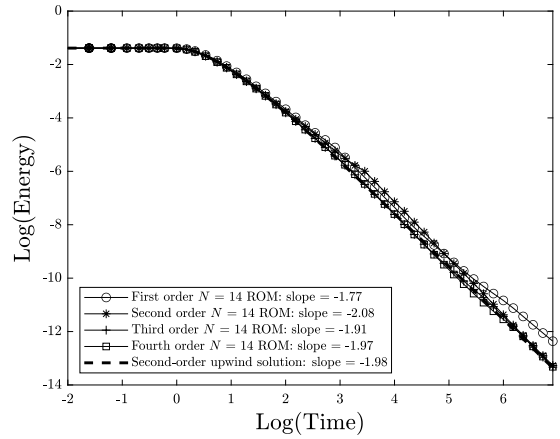}
\end{center}
\caption{The energy contained in the resolved modes of order $n=1,2,3,4$ for $N = 14$ ROMs of Burgers equation utilizing the optimal $\tau$ found for the fourth order model for up to time $t=1000$ depicted on a log-log plot. The slope was calculated using data in the time window $15 \leq t \leq 500$.}\label{fig:Burgers_long_energy_orders}
\end{figure}

\begin{figure}[h]
\centering
\includegraphics[width=0.6\textwidth]{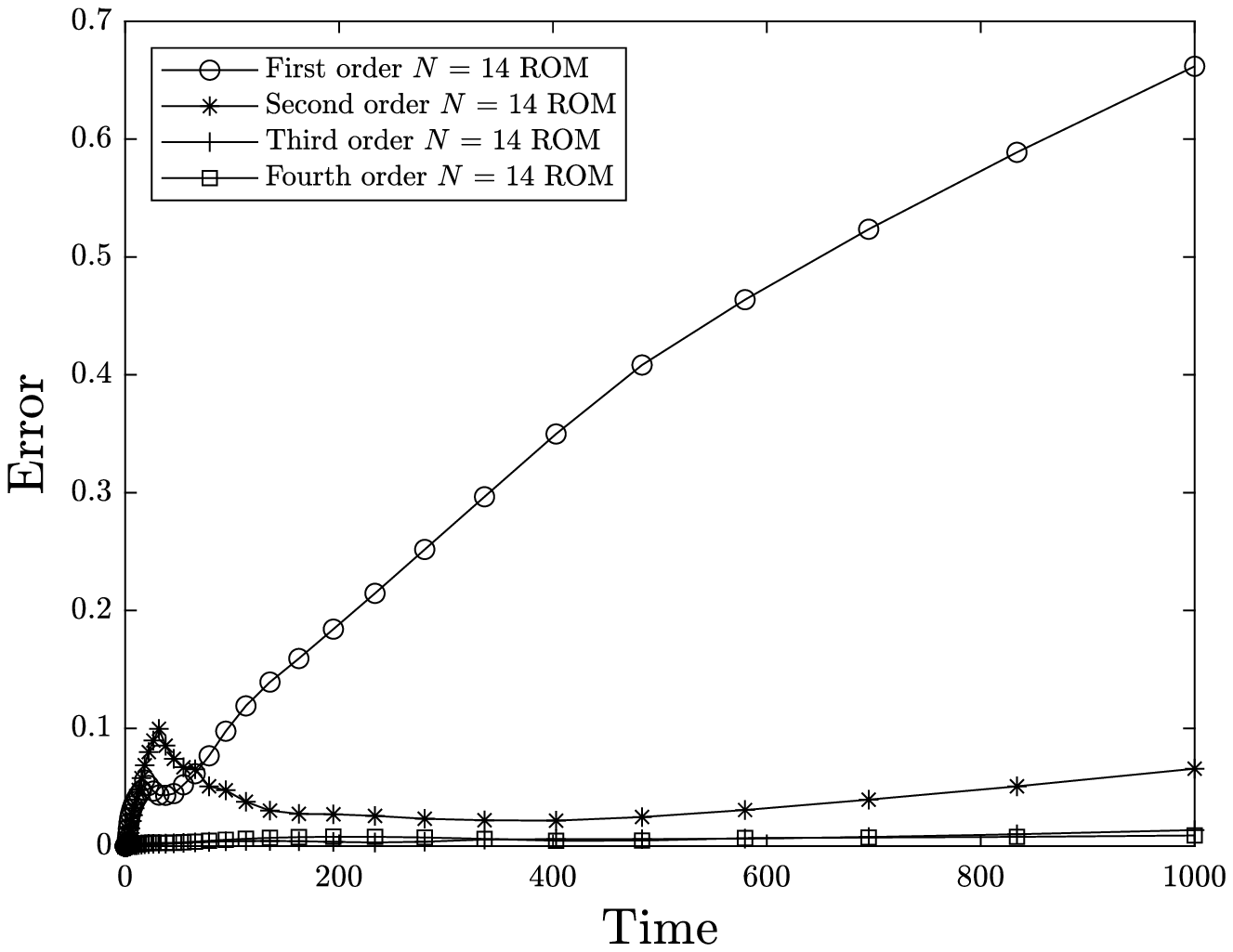} 
\caption{Relative error of the energy contained in the resolved modes of the $n=1,2,3,4$ order, $N = 14$ ROM solution of Burgers equation compared to a second-order upwind solution up to time $t$ = 1000. The relative error is calculated using $\frac{\sum_{k \in F} |u_k^{N,n}(t) - u_k(t)|^2}{\sum_{k \in F}  |u_k(t)|^2}$ where $u_k^{N,n}(t)$ is the solution for the ROM of size $N$ and order $n$, and $u_k(t)$ is the upwind solution.}\label{fig:Burgers_error_orders}
\end{figure}

\begin{figure}[h]
\centering
\includegraphics[width=0.6\textwidth]{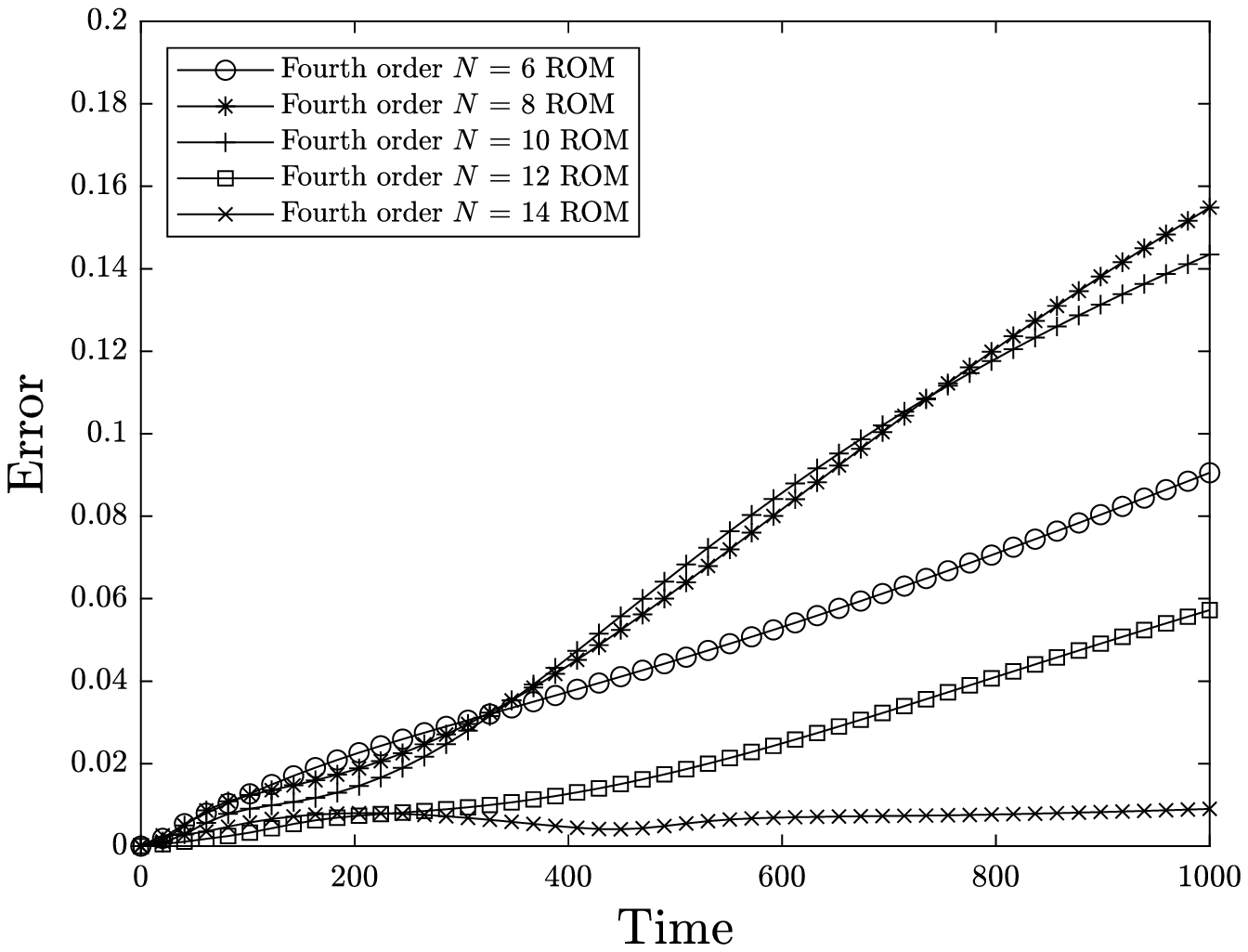} 
\caption{Relative error of the energy contained in the resolved modes of the fourth order ROM solution of Burgers equation compared to a second-order upwind solution up to time $t$ = 1000. The relative error is calculated using $\frac{\sum_{k \in F} |u_k^{N,4}(t) - u_k(t)|^2}{\sum_{k \in F}  |u_k(t)|^2}$ where $u_k^{N,4}(t)$ is the solution for the fourth order ROM of size $N$ and $u_k(t)$ is the upwind solution.}\label{fig:Burgers_error}
\end{figure}

 \begin{figure}[ht]
 \centering
 \subfigure[$t$ = 0.5]{{\includegraphics[width=0.45\textwidth]{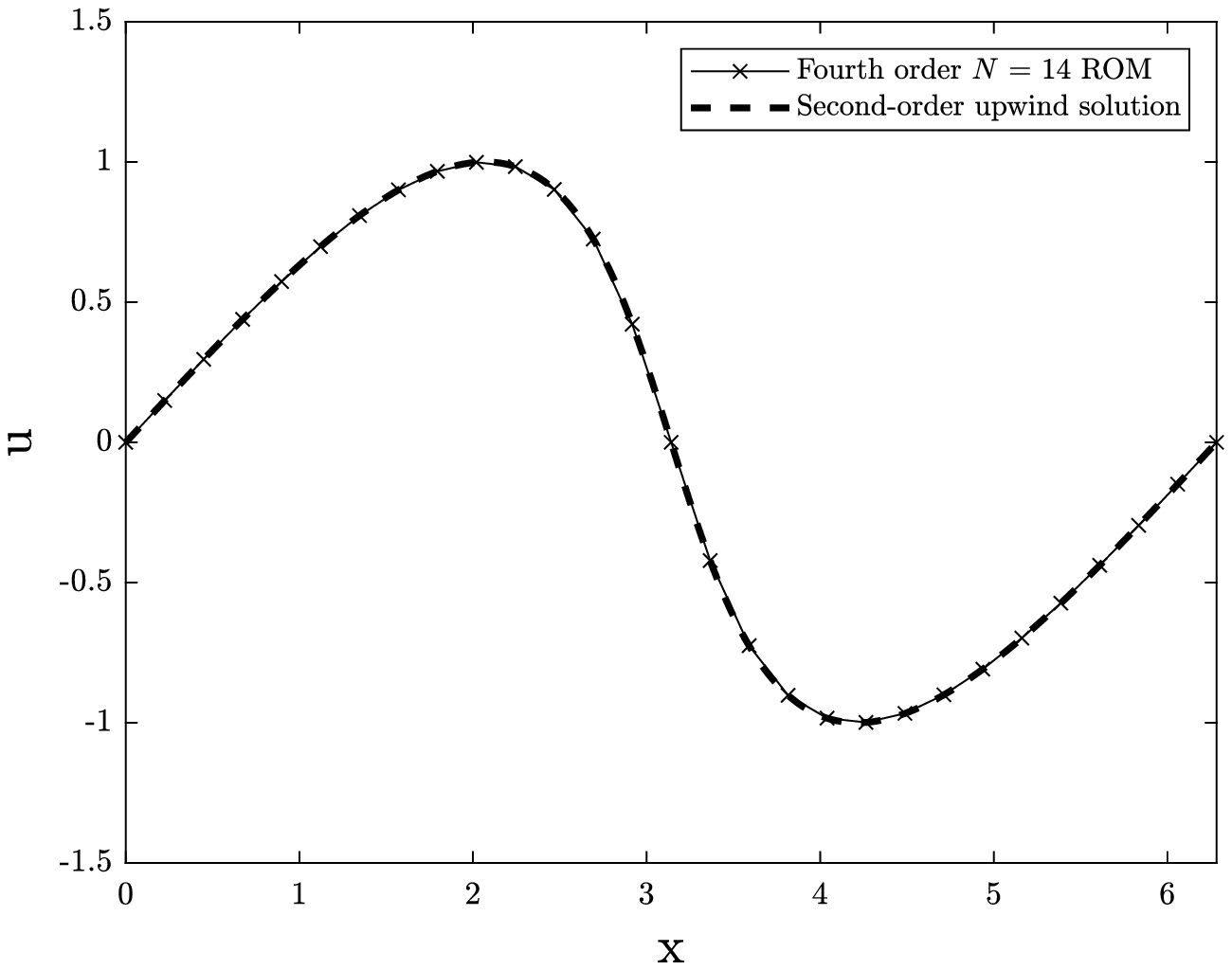} }}
 \qquad
 \subfigure[$t$ = 1.0]{{\includegraphics[width=0.45\textwidth]{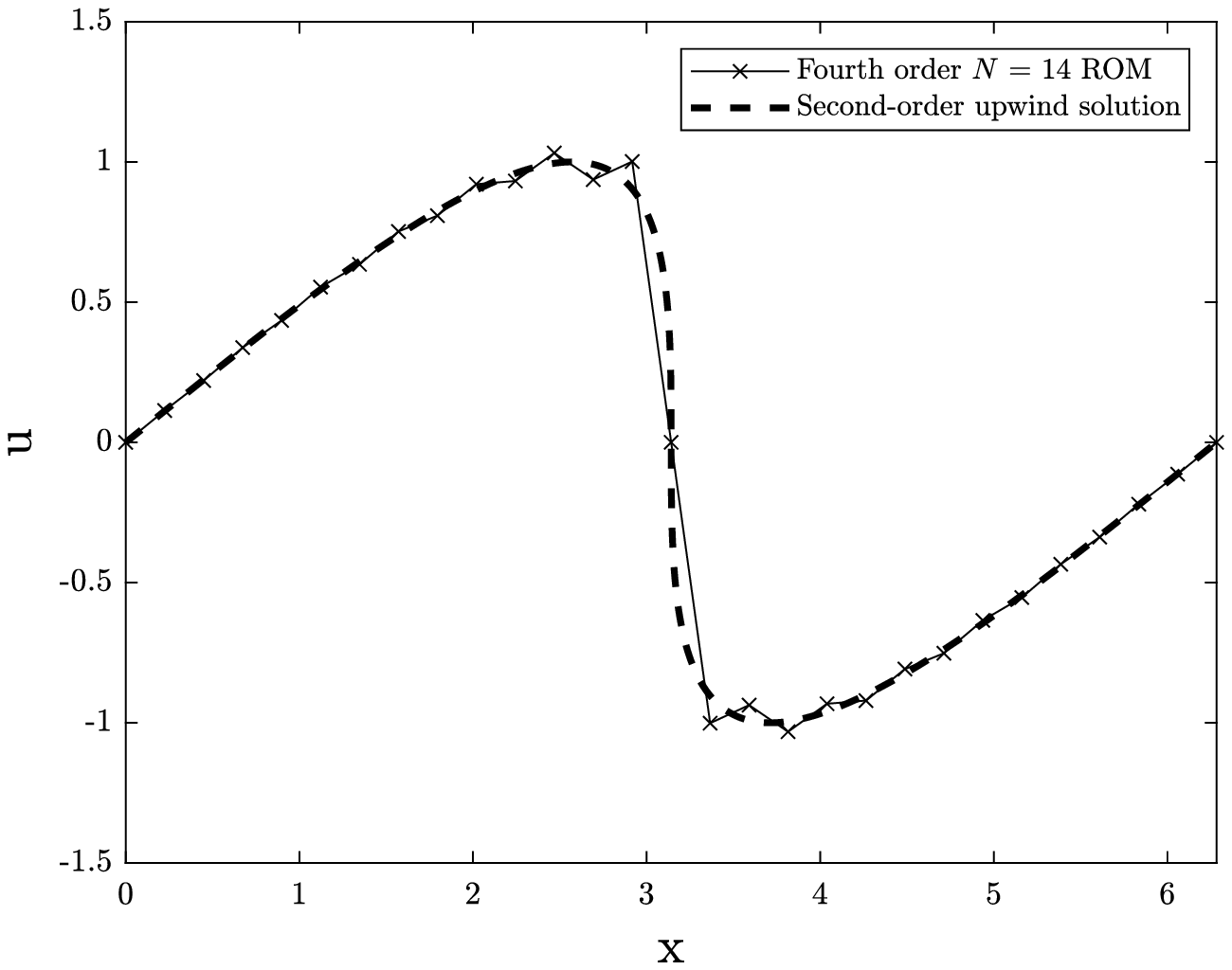} }}
 \qquad
 \subfigure[$t$ = 5.0]{{\includegraphics[width=0.45\textwidth]{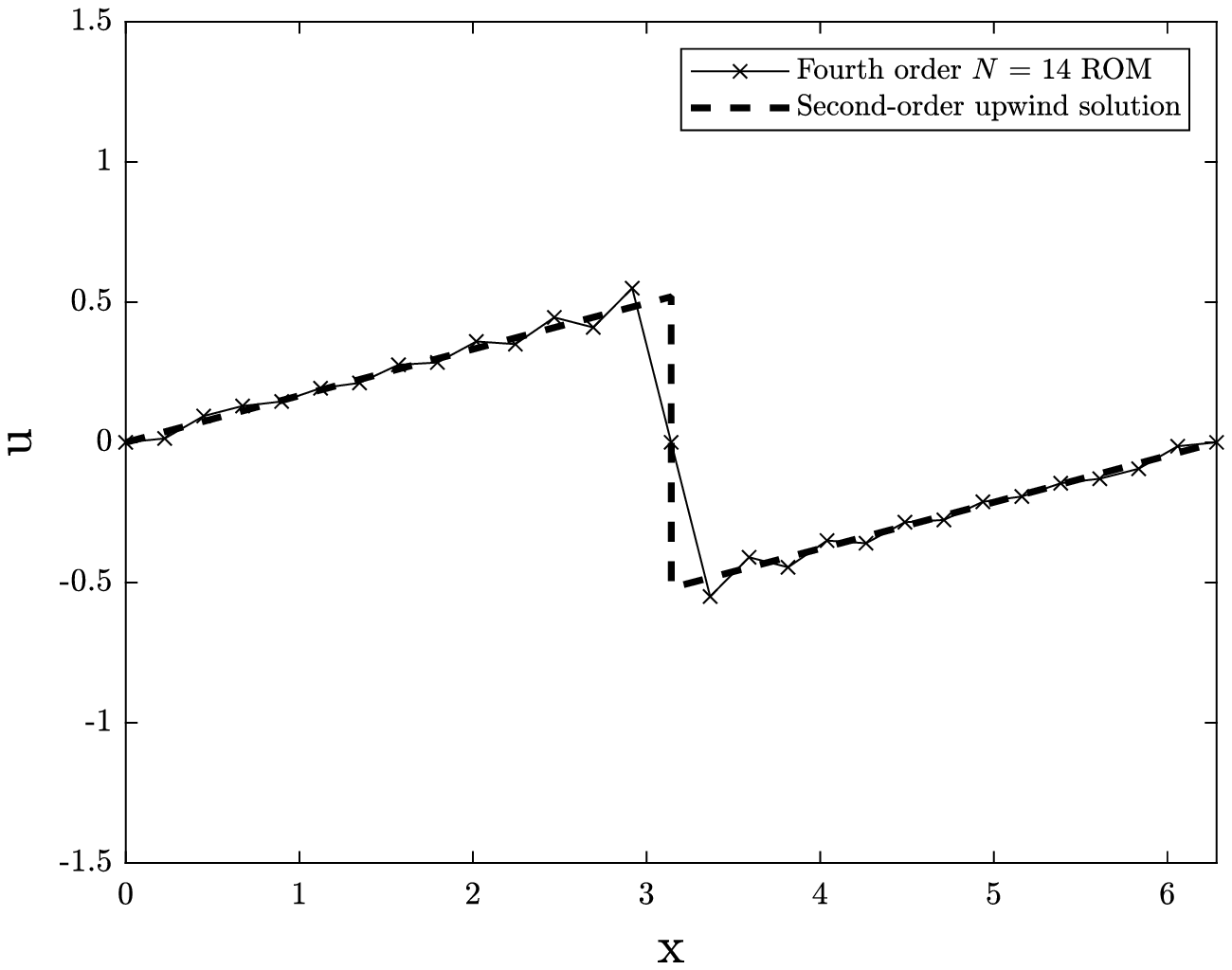} }}
 \qquad
 \subfigure[$t$ = 10.0]{{\includegraphics[width=0.45\textwidth]{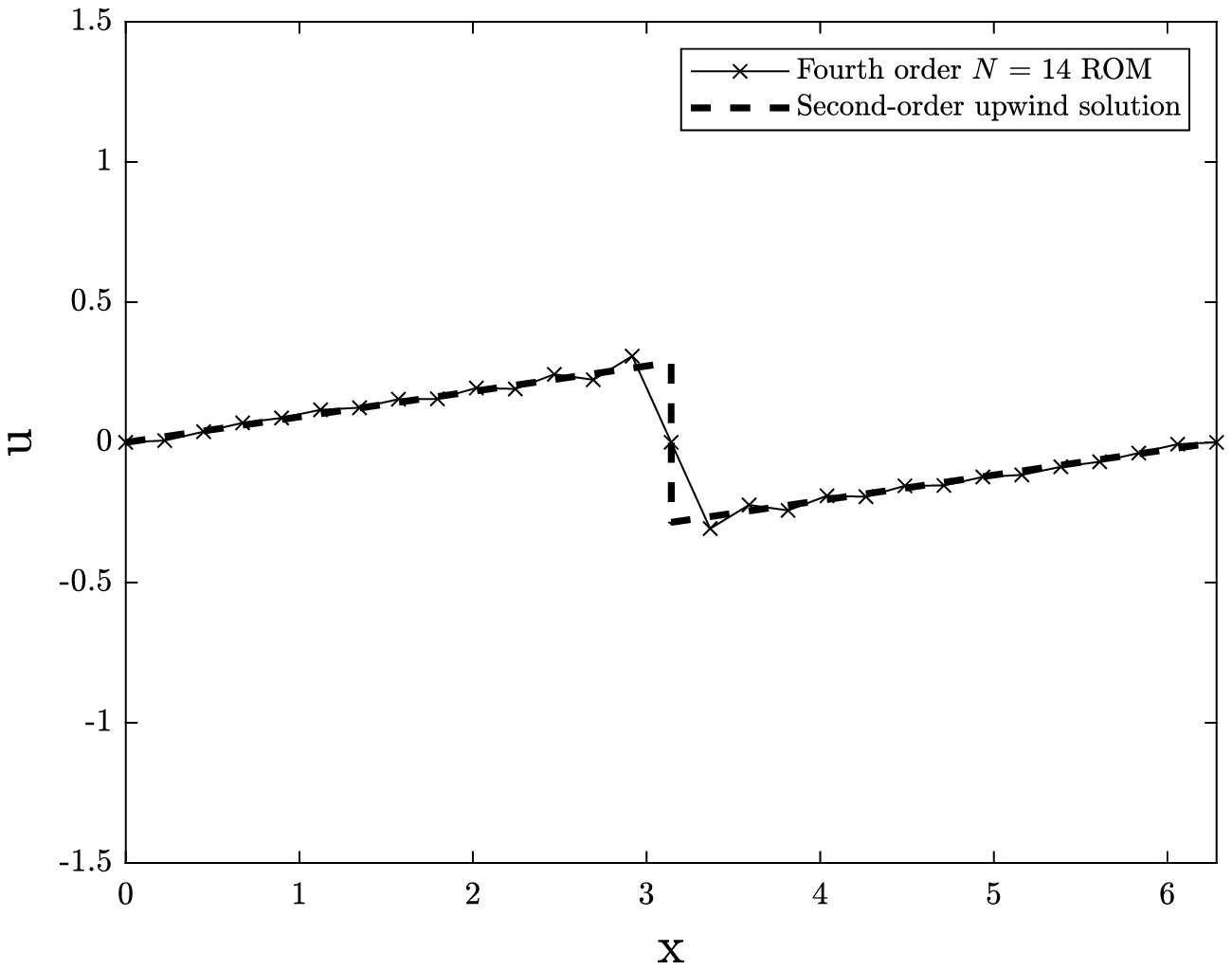} }}
 \caption{Fourth order N = 14 ROM real space solution of Burgers equation compared to a second-order upwind solution ($\Delta x = \frac{2\pi}{10000}$) up to time $t$ = 10.}\label{fig:Burgers_real_space}
 \end{figure}

\begin{figure}[h]
\centering
\includegraphics[width=0.6\textwidth]{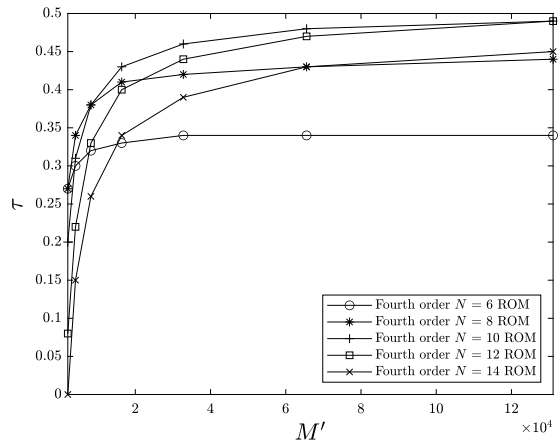} 
\caption{The $\tau$ predicted by the search procedure for $N = 6, 8,\dots, 14$ ROMs of Burgers equation as a function of full model size $M'$ for $M' = 2048$ to $M' = 131072$.}\label{fig:Burgers_asymptotic_tau}
\end{figure}

\begin{figure}[ht]
 \centering
 \subfigure{{\includegraphics[width=0.45\textwidth]{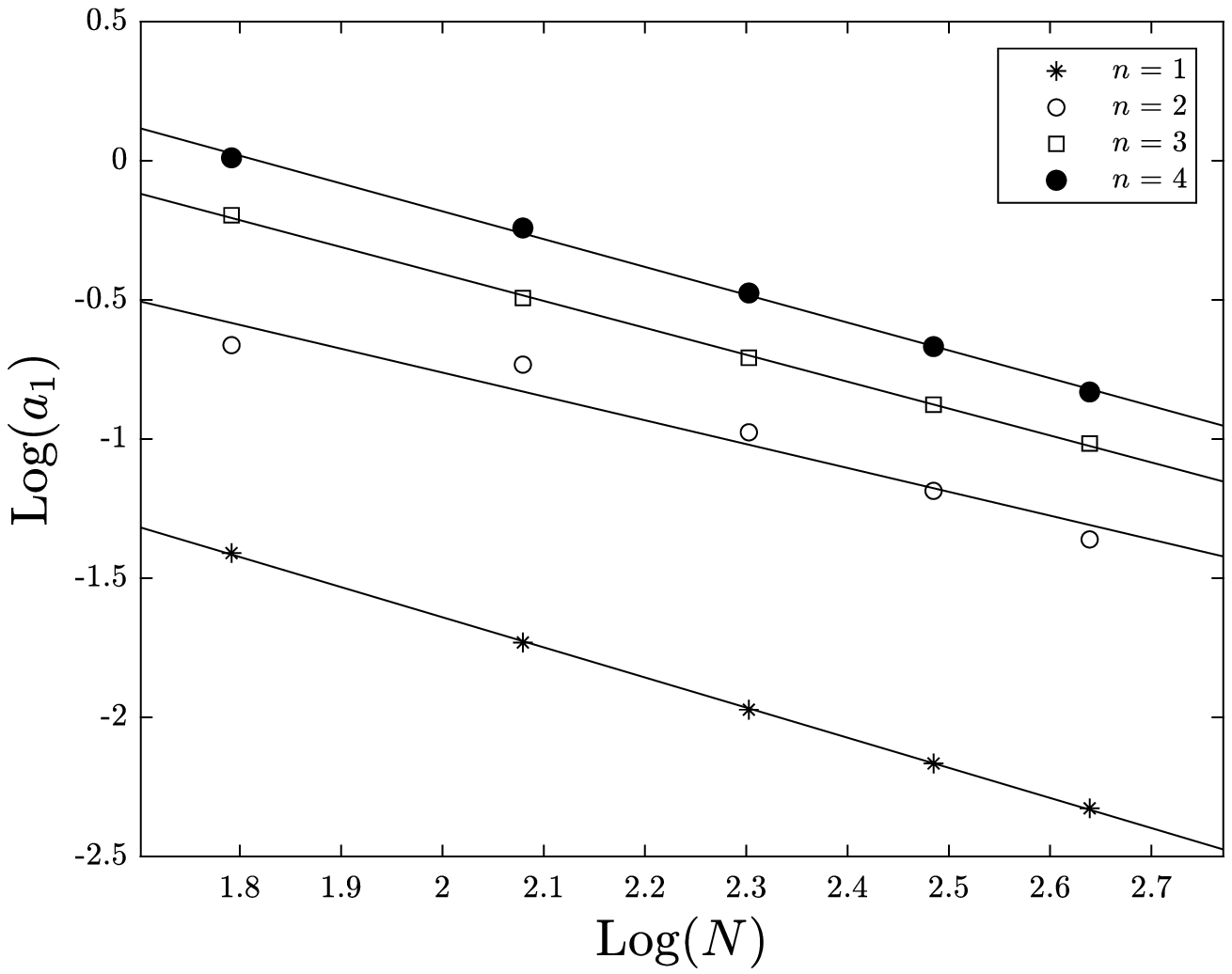} }}
 \qquad
 \subfigure{{\includegraphics[width=0.45\textwidth]{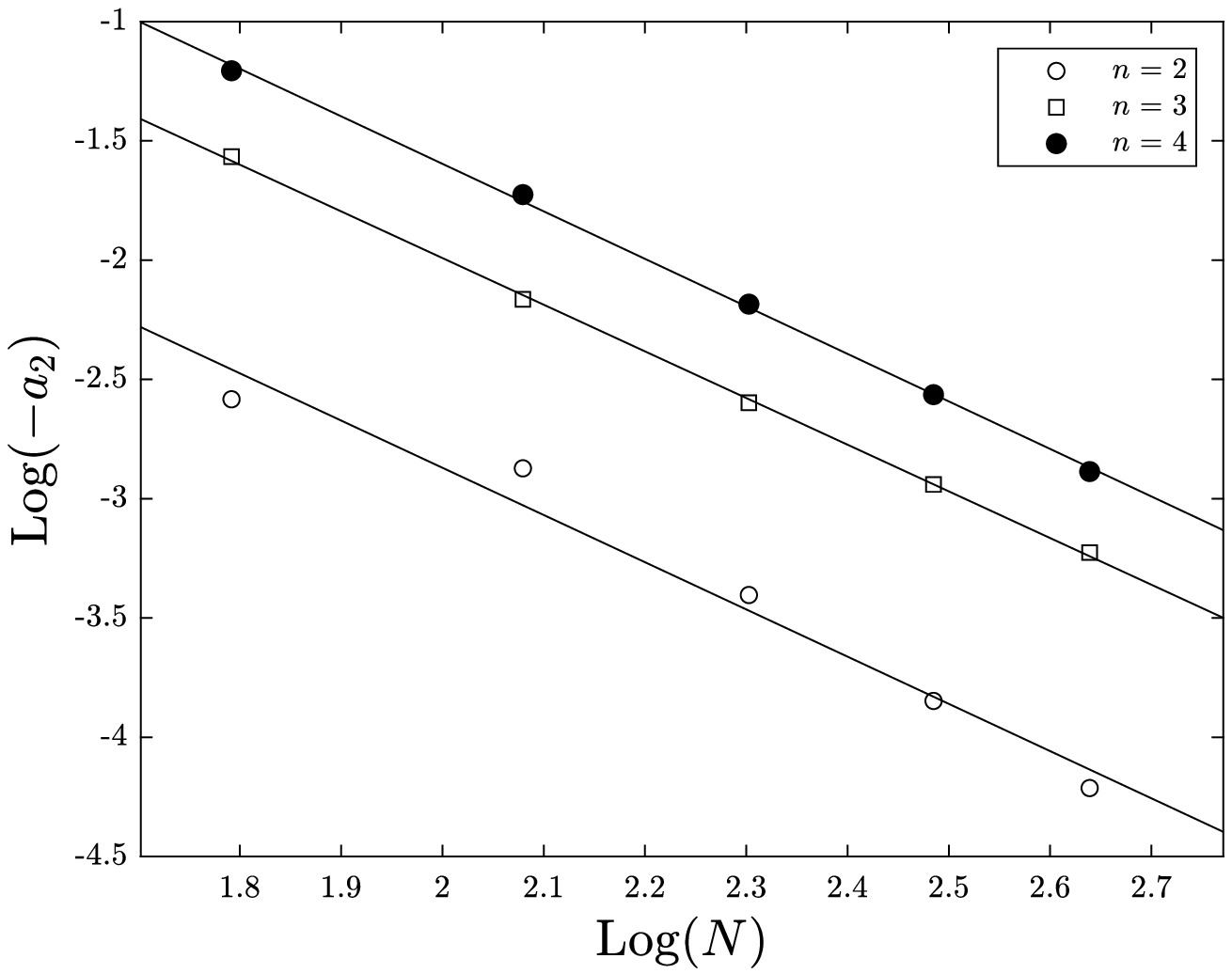} }}
 \qquad
 \subfigure{{\includegraphics[width=0.45\textwidth]{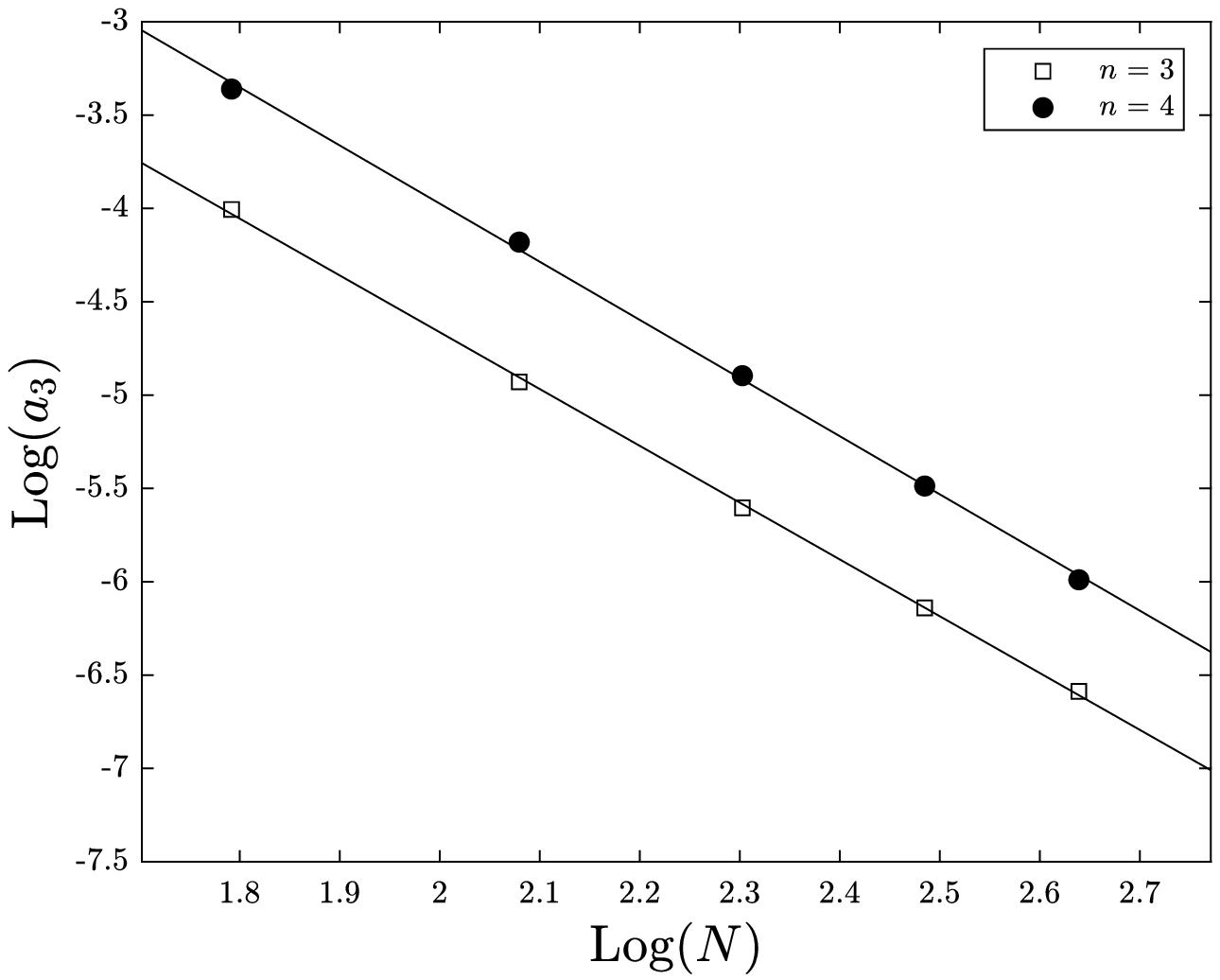} }}
 \qquad
 \subfigure{{\includegraphics[width=0.45\textwidth]{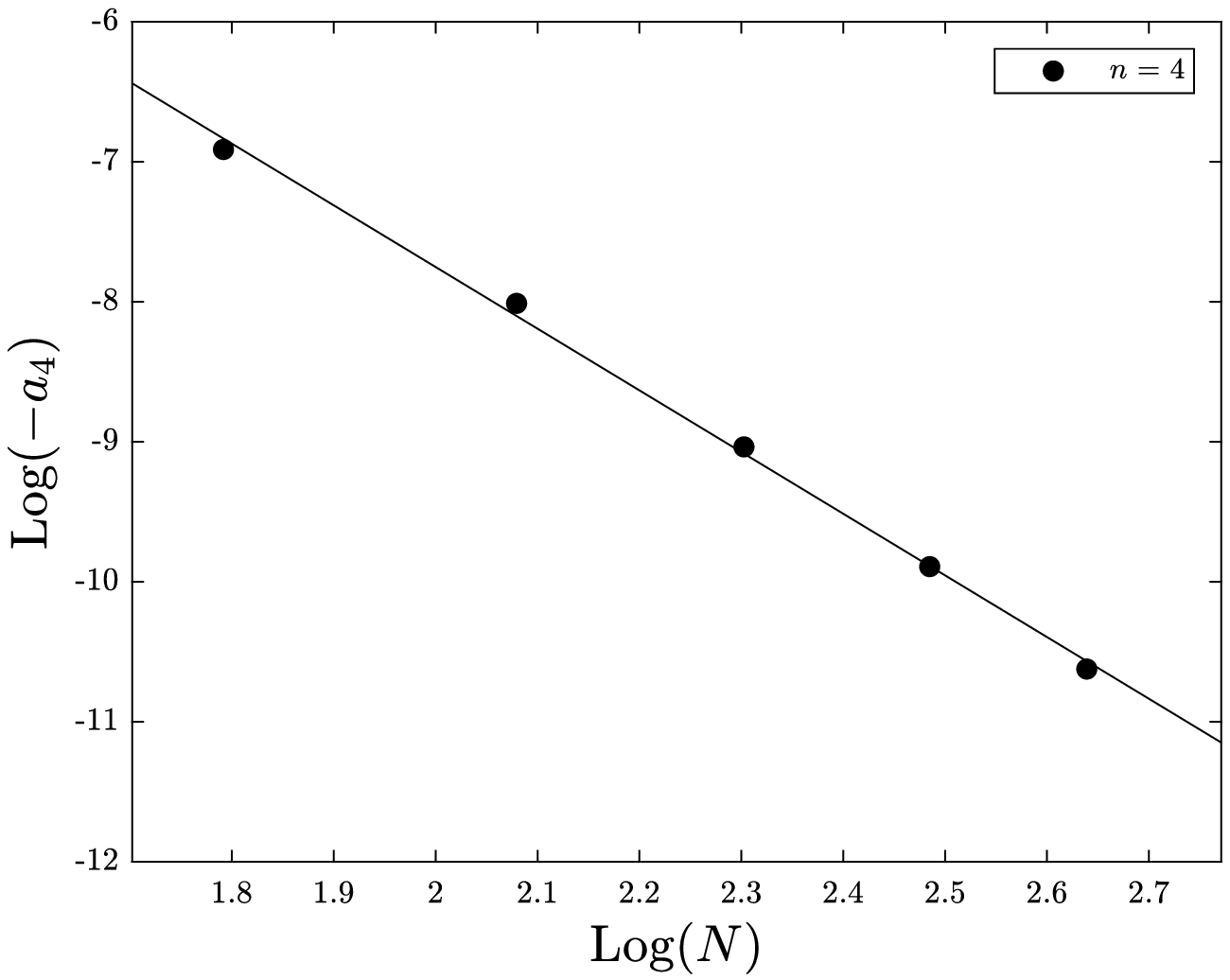} }}
 \caption{Optimal renormalization coefficients, $a_i$, for Burgers equation plotted versus $N$ on a log-log scale and utilizing $\tau = 0.4$.}\label{fig:Burgers_scaling_laws}
 \end{figure}

\begin{figure}[h]
\centering
\includegraphics[width=0.6\textwidth]{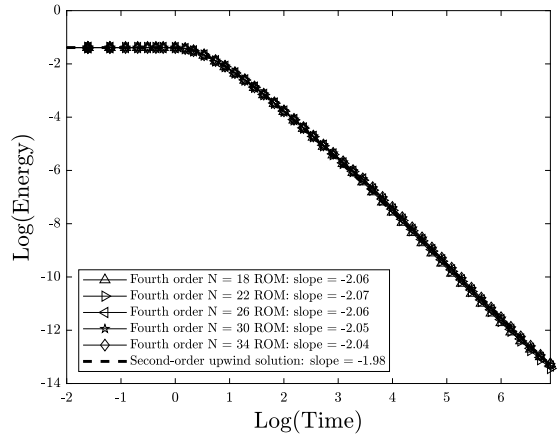} 
\caption{The energy contained in the resolved modes of order $n=4$ ROMs of Burgers equation utilizing the scaling laws presented and a value of $\tau = 0.4$ for up to time $t=1000$ depicted on a log-log plot. The slope was calculated using data in the time window $15 \leq t \leq 500$.}\label{fig:Burgers_long_energy_scaling}
\end{figure}

\begin{figure}[h]
\centering
\includegraphics[width=0.6\textwidth]{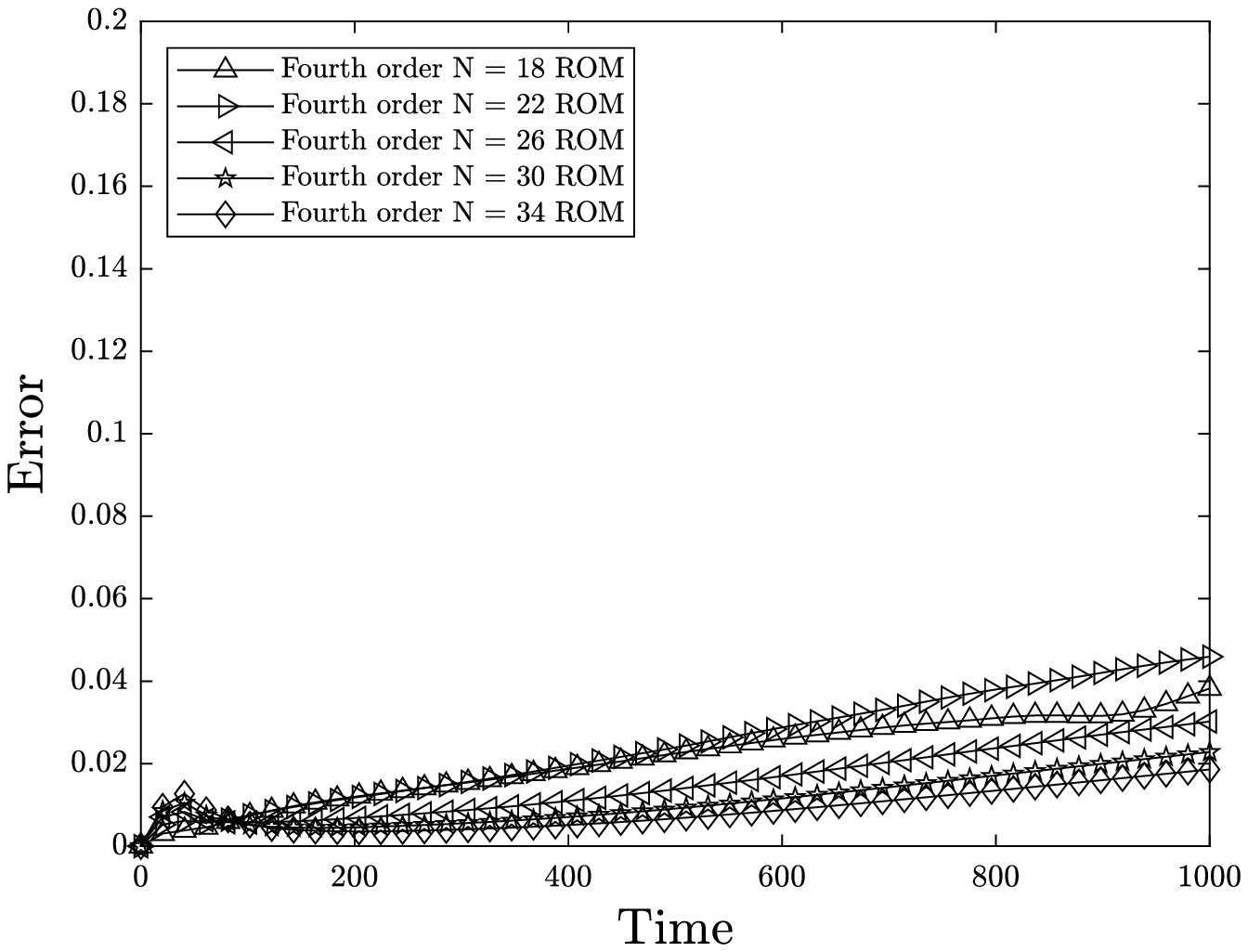} 
\caption{Relative error of the energy contained in the resolved modes of the fourth order ROM solution of Burgers equation utilizing the scaling laws presented and a value of $\tau = 0.4$ compared to a second-order upwind solution up to time $t$ = 1000. The relative error is calculated using $\frac{\sum_{k \in F} |u_k^{N,4}(t) - u_k(t)|^2}{\sum_{k \in F}  |u_k(t)|^2}$ where $u_k^{N,4}(t)$ is the solution for the fourth order ROM of size $N$ and $u_k(t)$ is the upwind solution.}\label{fig:Burgers_error_scaling}
\end{figure}

\begin{figure}[h]
\begin{center}
 \includegraphics[width=0.6\textwidth]{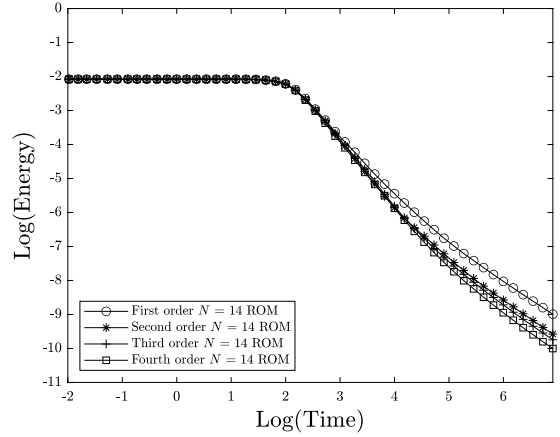}
\end{center}
\caption{The energy contained in the resolved modes of order $n=1,2,3,4$ for $N = 14$ ROMs of the 3D Euler equations utilizing $\tau = 1.0$ for up to time $t=1000$ depicted on a log-log plot.}\label{fig:Euler_long_energy_orders}
\end{figure}

\begin{figure}[h]
\centering
 \includegraphics[width=0.6\textwidth]{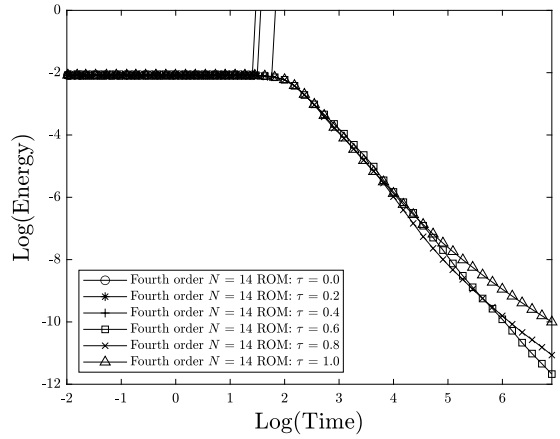}
\caption{The energy contained in the resolved modes of the N = 14 fourth order ROM of the 3D Euler equations and for $\tau$ = 0.0, 0.2, \dots, 1.0 for up to time $t=1000$ depicted on a log-log plot.}
\label{fig:Euler_long_energy_taus}
\end{figure}

\begin{figure}[ht]
 \centering
 \subfigure{{\includegraphics[width=0.45\textwidth]{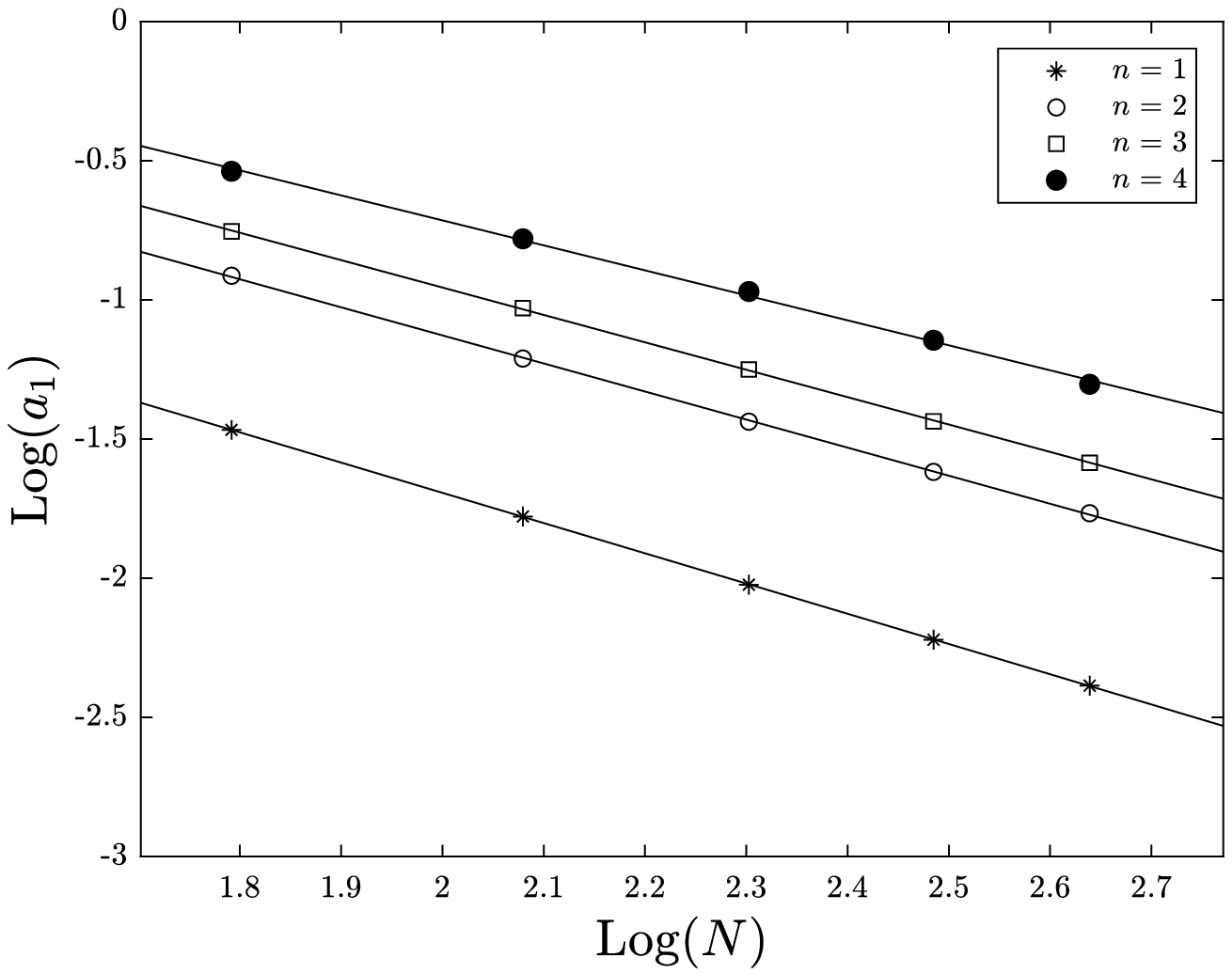} }}
 \qquad
 \subfigure{{\includegraphics[width=0.45\textwidth]{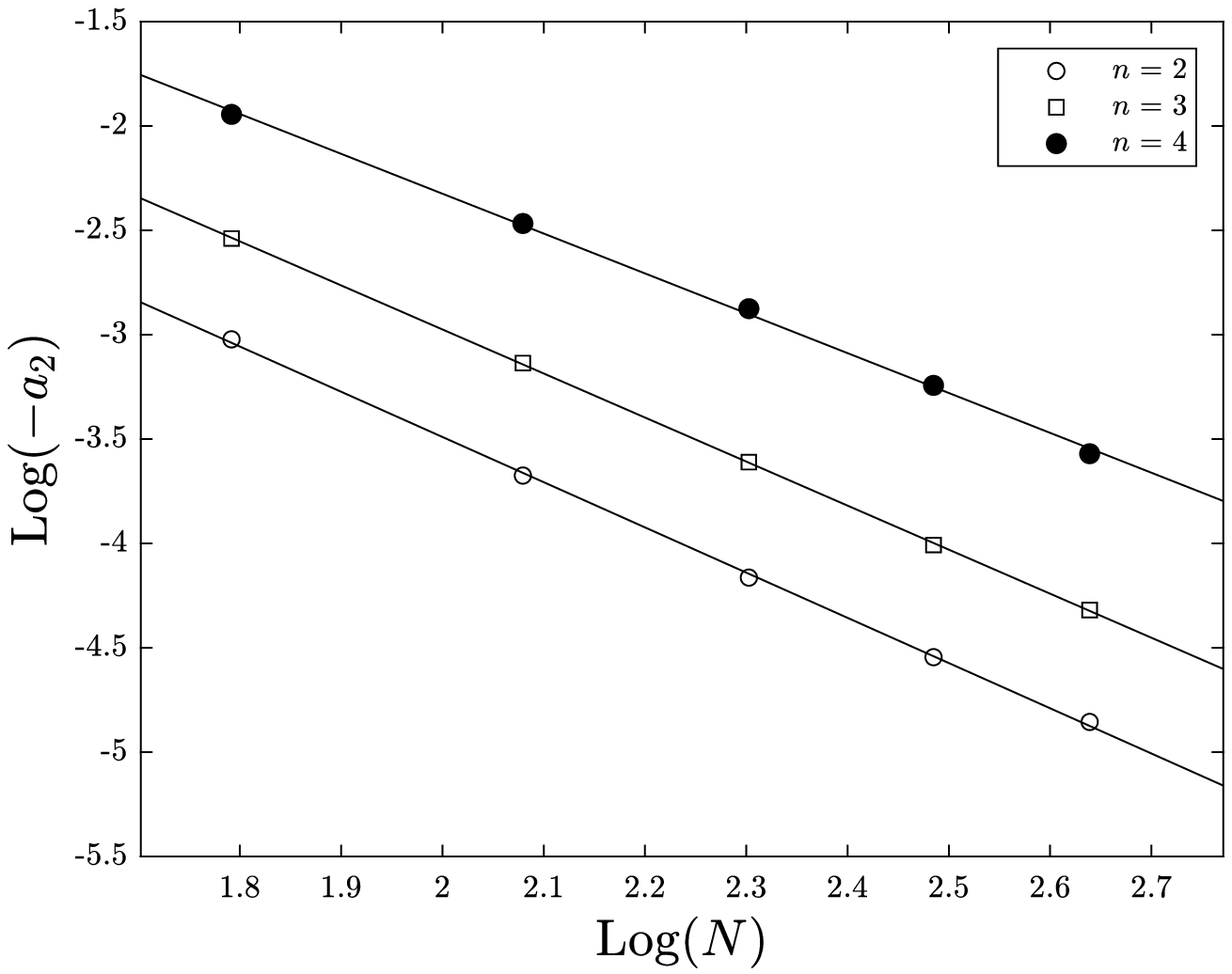} }}
 \qquad
 \subfigure{{\includegraphics[width=0.45\textwidth]{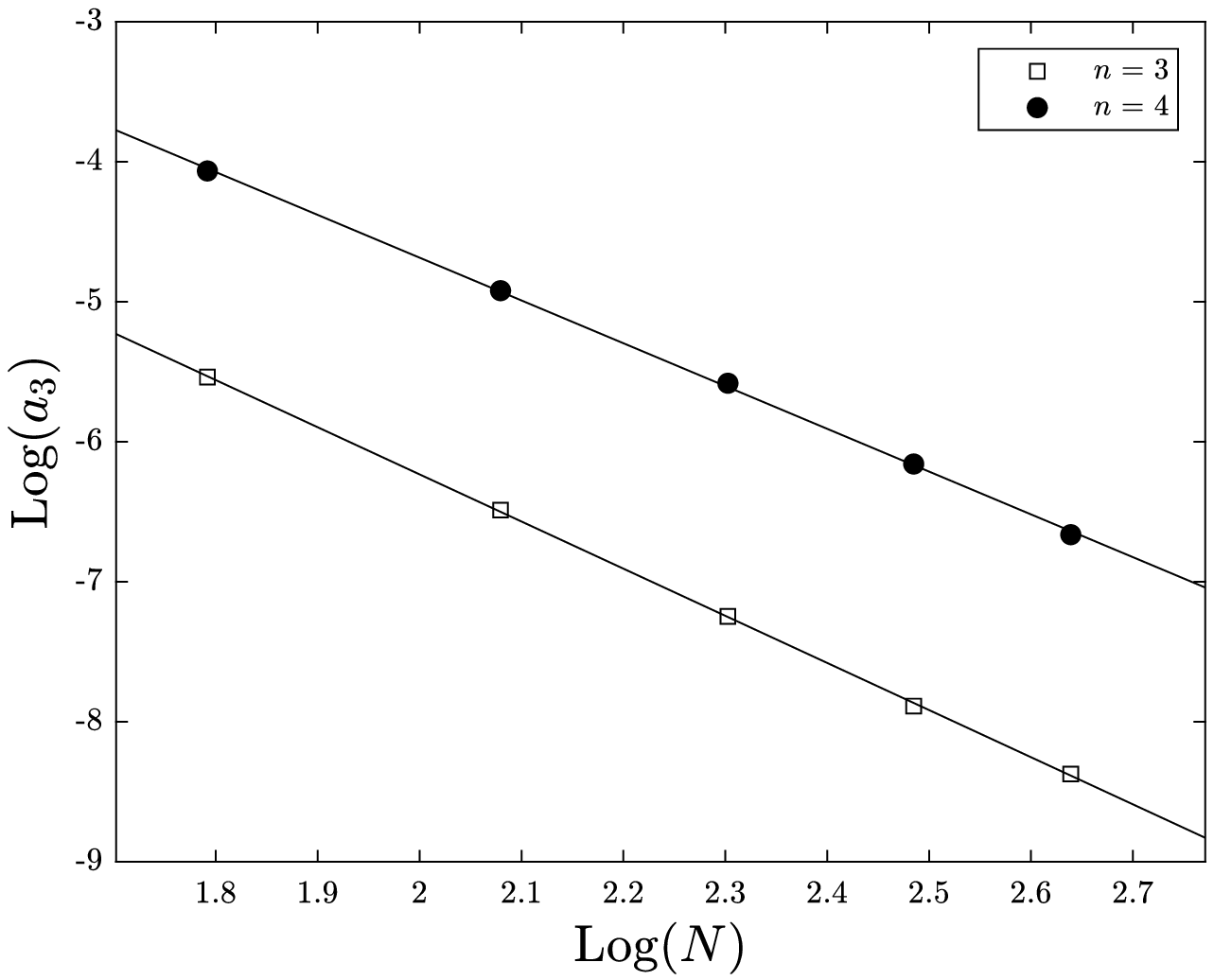} }}
 \qquad
 \subfigure{{\includegraphics[width=0.45\textwidth]{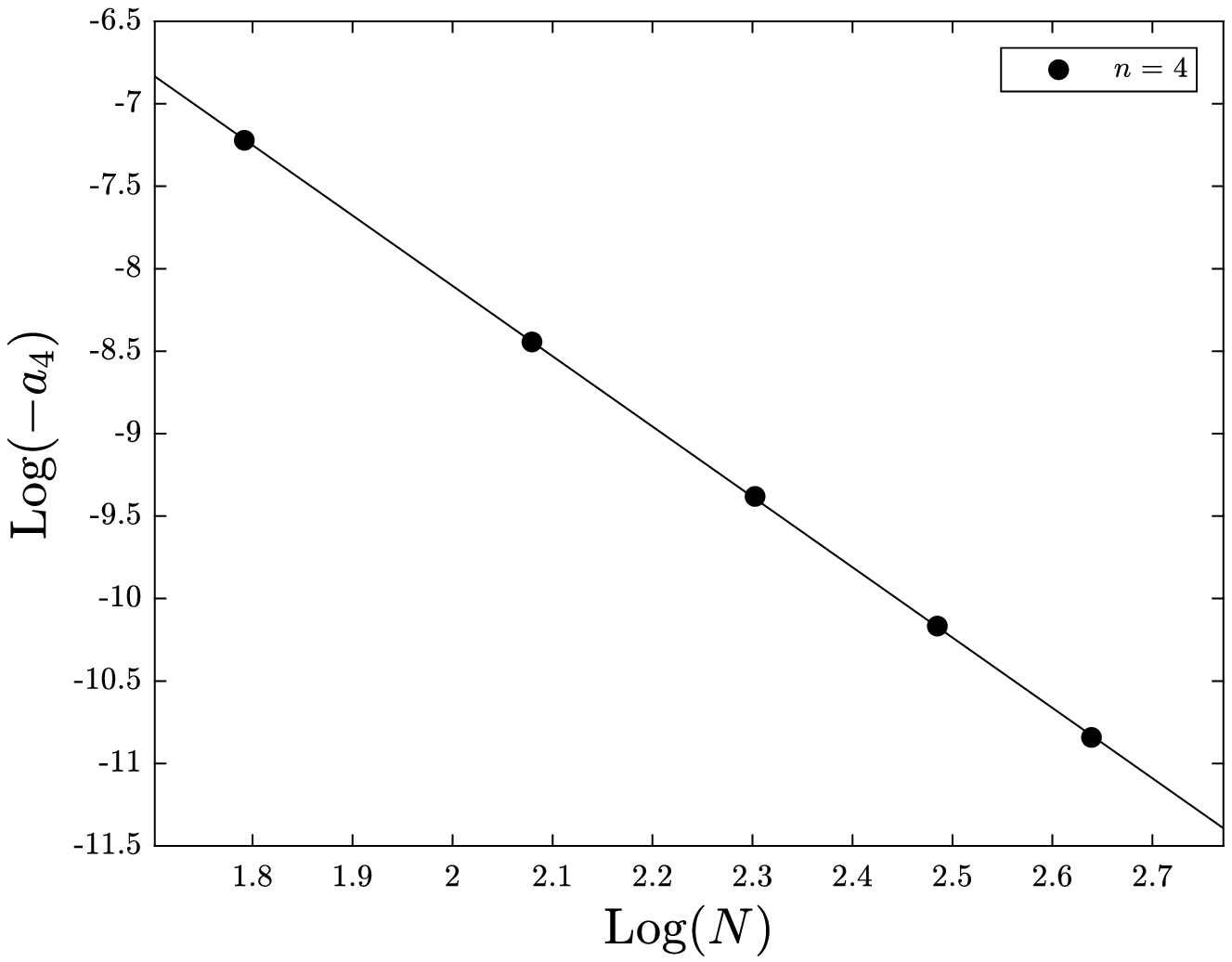} }}
 \caption{Optimal renormalization coefficients, $a_i$, for the 3D Euler equations plotted versus $N$ on a log-log scale and utilizing $\tau = 1.0$.}\label{fig:Euler_scaling_laws}
 \end{figure}

\begin{figure}[h]
\centering
 \includegraphics[width=0.6\textwidth]{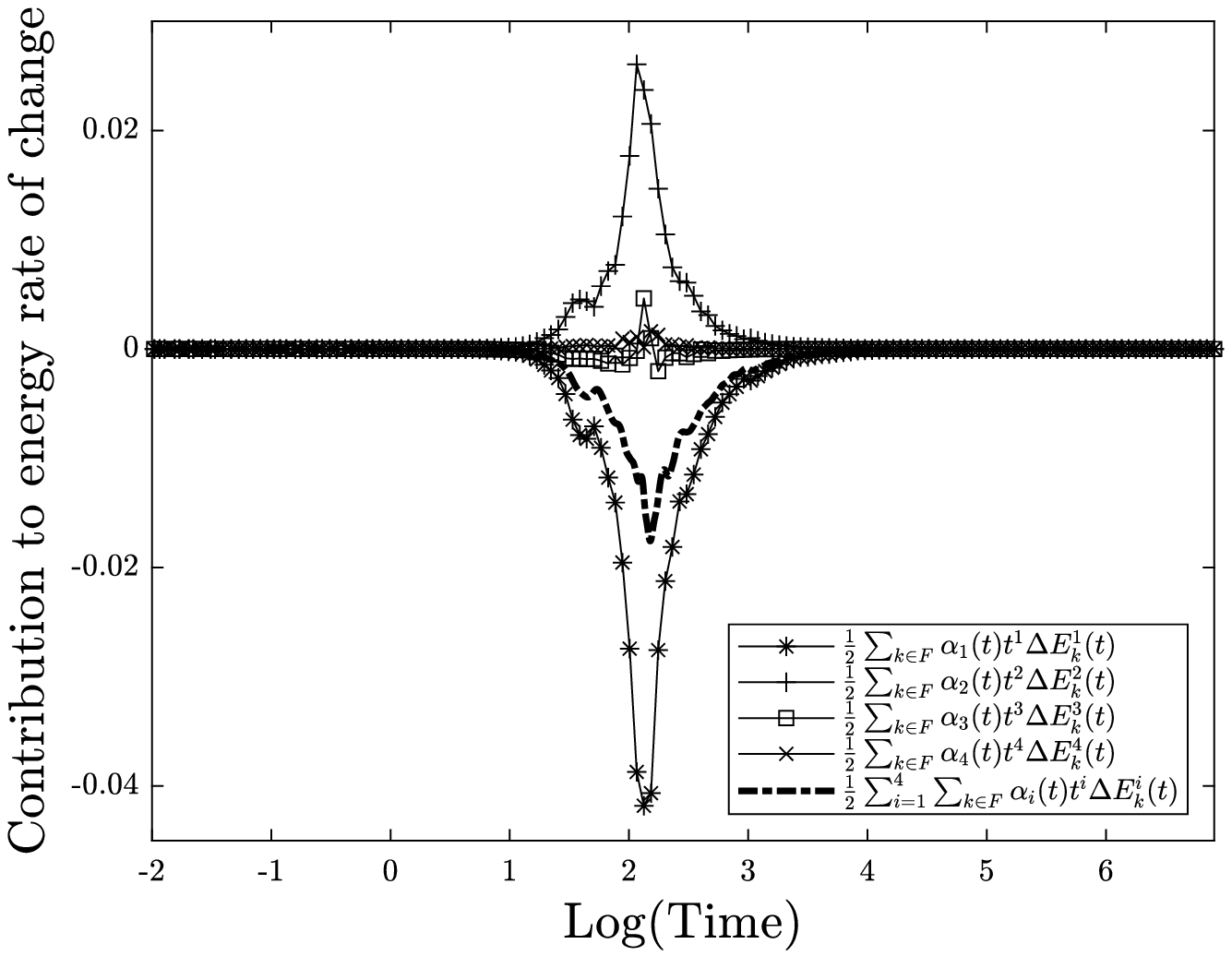}
\caption{The contribution of each order memory term to the rate of change of the energy in the resolved modes for the N = 14 fourth order ROM with $\tau=0.6.$}\label{fig:Euler_long_memory_tau_06}
\end{figure}

\begin{figure}[h]
\centering
\includegraphics[width=0.6\textwidth]{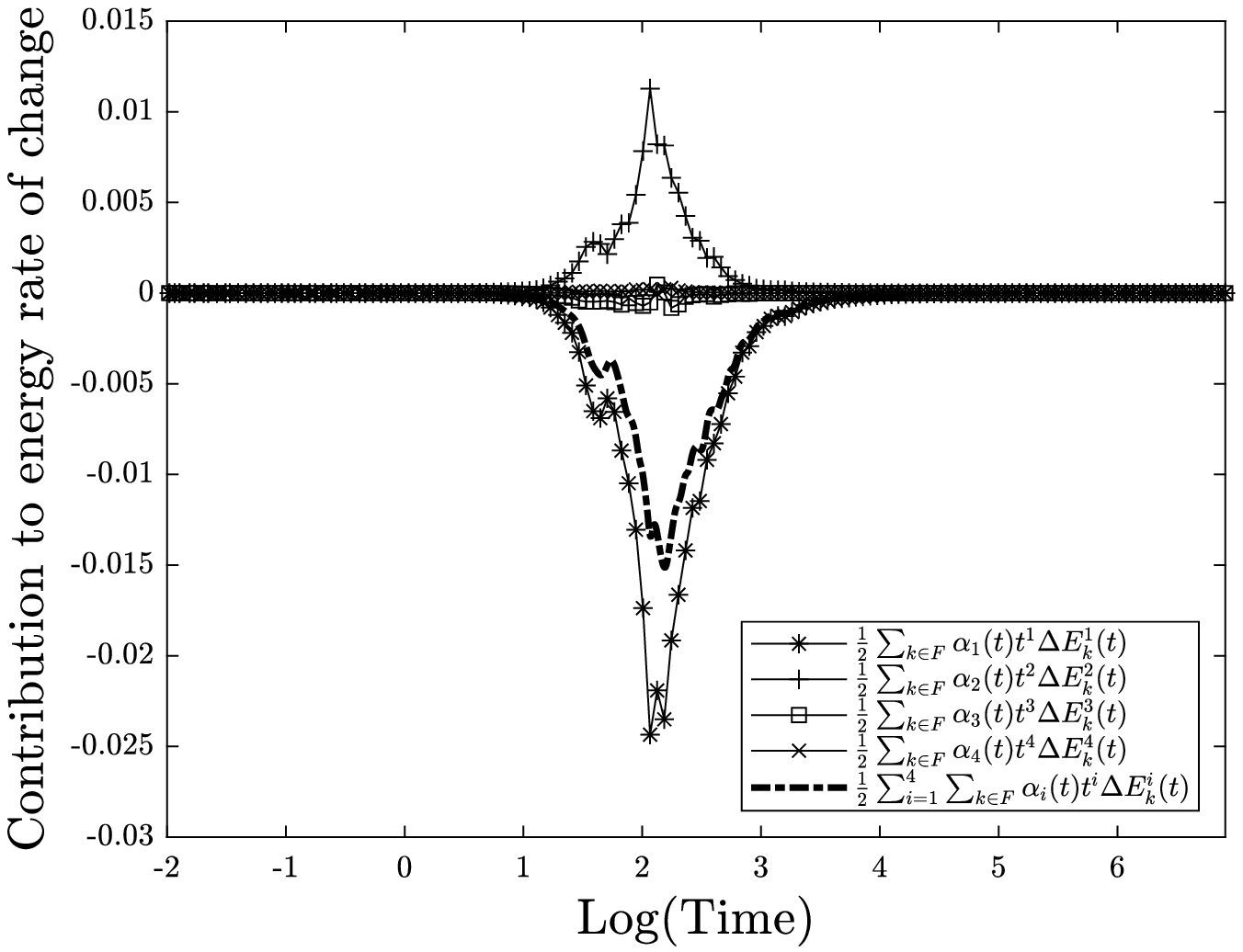}
\caption{The contribution of each order memory term to the rate of change of the energy in the resolved modes for the N = 14 fourth order ROM with $\tau=0.8.$}\label{fig:Euler_long_memory_tau_08}
\end{figure}

\begin{figure}[h]
\centering
\includegraphics[width=0.6\textwidth]{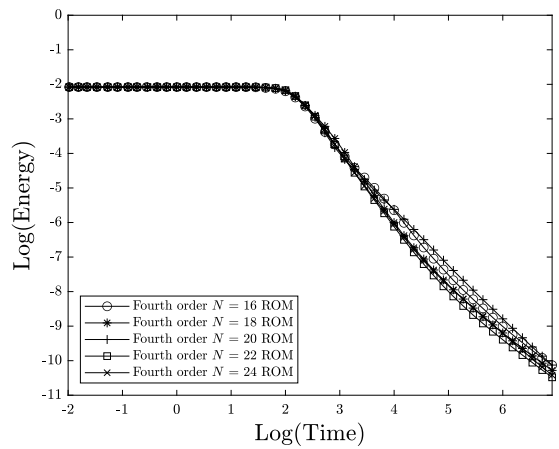} 
\caption{The energy contained in the resolved modes of order $n=4$ ROMs of the 3D Euler equations utilizing the scaling laws presented and a value of $\tau = 1.0$ for up to time $t=1000$ depicted on a log-log plot.}\label{fig:Euler_long_energy_scaling}
\end{figure}

\FloatBarrier

\begin{table}[h]
    \begin{center}
    \begin{adjustbox}{max width=0.48\textwidth}
    \begin{tabular}{|c||c|c|}
    \hline$\tau$ & Peak Time & Peak Value \\
    \hhline{|=#=|=|}
    0.6 & 8.8278 & -1.7577e-02 \\
    \hline
    0.8 & 8.9343 & -1.5129e-02  \\
    \hline
    1.0 & 9.1227 & -1.4066e-02  \\
    \hline
    \end{tabular}
    \end{adjustbox}
    \caption{Peak time and corresponding peak value of the total rate of change of the energy in the resolved modes for the N = 14 fourth order ROM for $\tau = 0.6,0.8,1.0$.} 
    \label{tab:Euler_memory_peak}
    \end{center}
\end{table}

\FloatBarrier

\end{document}